\documentclass[apj]{emulateapj}

\usepackage{calc}

\shorttitle{SN~2008ax Revisited}
\shortauthors{Folatelli et al.}

\begin{document}

\title{The Progenitor of the Type IIb SN~2008ax Revisited}

\author{%
  Gast\'on Folatelli\altaffilmark{1,2,3},
  Melina C.\ Bersten\altaffilmark{1,2,3},
  Hanindyo Kuncarayakti\altaffilmark{4,5},
  Omar G.\ Benvenuto\altaffilmark{1,2},
  Keiichi Maeda\altaffilmark{6,3},
  and
  Ken'ichi Nomoto\altaffilmark{3,7},
}
\altaffiltext{1}{Instituto de Astrof\'isica de La Plata, Argentina}
\altaffiltext{2}{Facultad de Ciencias Astron\'omicas y
  Geof\'{\i}sicas, Universidad Nacional de La Plata, Paseo del Bosque
  S/N, B1900FWA La Plata, Argentina}
\altaffiltext{3}{Kavli Institute for the Physics and Mathematics of
  the Universe (WPI), The University of Tokyo, Kashiwa, Chiba
  277-8583, Japan}
\altaffiltext{4}{Millennium Institute of Astrophysics (MAS), Casilla
  36-D, Santiago, Chile}  
\altaffiltext{5}{Departamento de Astronom\'ia, Universidad de Chile,
  Casilla 36-D, Santiago, Chile}  
\altaffiltext{6}{Department of Astronomy, Kyoto University,
  Kitashirakawa-Oiwake-cho, Sakyo-ku, Kyoto 606-8502, Japan} 
\altaffiltext{7}{Hamamatsu Professor}

\email{gaston@fcaglp.unlp.edu.ar}

\begin{abstract}
\noindent {\em Hubble Space Telescope} observations of the site of
the supernova (SN)~2008ax obtained in 2011 and 2013 reveal that the possible
progenitor object detected in pre-explosion images was in fact
multiple. Four point sources are resolved in the new,
higher-resolution images. We identify one of the sources with the
fading SN. The other three objects are consistent with single
supergiant stars. We conclude that their light contaminated the
previously identified progenitor candidate. After subtraction of
these stars, the progenitor appears to be significantly fainter and
bluer than previously measured. Post-explosion photometry at the SN
location indicates that the progenitor object has disappeared. If
single, the progenitor is compatible with a supergiant star of B to
mid-A spectral type, while a Wolf-Rayet (WR) star would be too
luminous in the ultraviolet to account for the observations. Moreover,
our hydrodynamical modelling shows the pre-explosion mass was 4--5
$M_\odot$ and the radius was 30--50 $R_\odot$, which is incompatible
with a WR progenitor. We present a possible 
interacting binary progenitor computed with our evolutionary
models that reproduces all the observational evidence. A companion
star as luminous as an O9-B0 main-sequence star may have remained
after the explosion.
\end{abstract}


\section{INTRODUCTION}
\label{sec:intro}

\noindent The question of what stellar system produce the different
types of core-collapse supernovae (SNe) is still open. While it is
accepted that hydrogen-rich (Type II-Plateau) SNe arise from relatively
low-mass stars in the red supergiant phase \citep{Smartt09}, the
nature of hydrogen-deficient (Types Ib and Ic) SNe progenitors remains
unclear \citep{Eldridge13}. One crucial aspect is understanding what
mechanisms are at play for envelope removal. In this respect it is
interesting to focus the attention on the intermediate group of Type
IIb SNe (SNe~IIb) whose spectra evolve from hydrogen rich to hydrogen
poor and which has been suggested to originate from stars that lost
most of their H-rich envelopes. Proposed mass-loss mechanisms are
those of strong stellar winds or interaction in close binary
systems. The former mechanism requires a very large mass of the SN
progenitor, possibly with a Wolf-Rayet (WR) or luminous blue variable
(LBV) nature. 

Remarkably, the firmest identifications of SN-IIb progenitors in
pre-explosion images---i.e., those of SNe~1993J and 2011dh---have been
associated to moderate-mass supergiants of spectral types F to K
\citep{Aldering94,VanDyk02,Maund11,VanDyk11},
hotter than the progenitors of Type II-Plateau SNe. Moreover, they
were suggested to have formed close binary systems by the possible
detection of their companion stars \citep{Maund04,Fox14,Folatelli14}. 
SNe~2008ax and 2013df are the other two SNe~IIb with possible
progenitor identifications in pre-explosion imaging. The case of
SN~2008ax is revised in this work using newly available data. For
SN~2013df, the candidate is again a cool, K-type supergiant star
\citep{VanDyk14}. These cases appear to indicate a 
prevailing close binary origin for SNe~IIb. However, \citet{Gal-Yam14}
found signatures of a WR-type wind in very ealy spectra of the SN~IIb
2013cu. 

The SN~IIb 2008ax has been an important case of study due to its small
distance and the availability of pre-explosion {\em Hubble Space
  Telescope} ({\em HST}) imaging and extensive multi-wavelength 
followup. \citet{Crockett08} studied the pre-SN images and found a
source at the SN location that 
was possibly the progenitor. With the then available data, the authors
selected two out of a series of possible scenarios: a WR star that
retained a small amount of hydrogen, or an interacting binary
system. The light curves and spectra indicated a relatively low
ejected mass and normal kinetic energy
\citep{Pastorello08,Roming09,Tsvetkov09,Maurer10,Taubenberger11,Jerkstrand15}. 
The low mass was generally interpreted as an indication against a single
massive WR progenitor. However, by including rotation in the
evolutionary models of massive stars \citet{Georgy12} found a suitable
progenitor born with 20 $M_\odot$ and ending with the correct core mass,
hydrogen content, luminosity and color to explain the complete set of
observations. \citet{Groh13a} reinterpreted the final stage of the
rotating model as a luminous blue variable (LBV) star and suggested that
LBVs may be the progenitors of some core-collapse SNe.

One intriguing aspect of SNe~IIb is the possible relation between
progenitor radius (or H-envelope mass) and circumstellar medium
(CSM) density. If the progenitor star can be detected in pre-SN
observations, a radius can be determined. Alternatively, the shape of
the UV-optical light curve during the first few days after explosion
can also indicate the extent of the exploding star depending on the
strength of the cooling emission from the shocked envelope
\citep{Woosley94,Bersten12,Bersten14}. The CSM can be probed through
radio and X-ray observations and this can be linked to the mass-loss
history of the progenitor. When a dense CSM is present shock-heated
gas can also produce optical emission at late times, typically evident
in box-like emission lines in the spectra and a slow-down of the light
curve as compared to the usual radioactive-power decline rate. These
phenomena were observed in the SNe~IIb 1993J
\citep{Matheson00a,Matheson00b} and 2013df \citep{Maeda15}. 

\citet{Chevalier10} suggested that radio properties
of core-collapse SNe provide information about the size of the
progenitor and proposed that SNe~IIb were divided into those with
compact (WR-like) and extended (supergiant) progenitors. SN~2008ax was
placed in the compact category while SN~1993J belonged to the extended one. 
\citet{Bersten12}, however, showed that such a connection breaks down
for SN~2011dh. Nonetheless, by restricting to the three SNe~IIb with
direct progenitor detections and estimates of the CSM density
(SNe~1993J, 2011dh, and 2013df),
\citet{Maeda15} suggest that SNe with more extended progenitors may
have denser surrounding media, possibly due to a larger mass-loss rate
near the time of explosion. The range of radii goes from $\approx$200
$R_\odot$ in SN~2011dh to $\approx$600 $R_\odot$ in SNe~1993J and
2013df. There is a wide variety of SNe-IIb light curves observed
during the first few days after explosion, which may indicate a
range of progenitor radii. However, it is not clear whether the relation
suggested by \citet{Maeda15} holds in general, and if it represents a
continuum of progenitor properties or two distinct groups. Adding more
objects to the CSM density--progenitor radius diagram is thus of
critical importance to understand the evolutionary paths and mass-loss
history of massive stars. 

Since the work by \citet{Crockett08}, new {\em HST} imaging of the
field of SN~2008ax was obtained with longer exposure times and better spatial
resolution than the pre-explosion ones. The new images allowed us to
revisit the SN site in order to study the nature of the progenitor. In
Section~\ref{sec:obs} we present the high spatial resolution images 
used in this work and provide estimates for the distance and
extinction toward SN~2008ax. Section~\ref{sec:ana} is dedicated to the 
analysis of the pre- and post-explosion photometry at the SN
site. Section~\ref{sec:prog} presents the possible properties of the
progenitor object based on hydrodynamical and stellar evolution
models. Our conclusions are given in 
Section~\ref{sec:concl}. 

\section{OBSERVATIONAL DATA}
\label{sec:obs}

\subsection{High-Resolution Imaging}
\label{sec:img}

\noindent This work is based on high-resolution images of the
field where SN~2008ax went off in NGC~4490. The observations are
summarized in Table~\ref{tab:obs}, including the images previously
analyzed by \citet{Crockett08} and new imaging obtained in 2011 and
2013. The former dataset included images in four optical bands
obtained at three different epochs with the Wide Field Planetary
Camera 2 (WFPC2) mounted on {\em HST}. The SN site landed on WF
detectors for the $F450W$, $F606W$, and $F814W$ 
frames, yielding a pixel scale of $0{\farcs}1$ pixel$^{-1}$. In the
$F300W$ frames the SN site was on the PC chip, with a pixel scale of
$0{\farcs}046$ pixel$^{-1}$. The new imaging was obtained in two
epochs (in July 2011 and October 2013) using the Wide Field Camera 3
(WFC3) of {\em HST} with a pixel scale of $0{\farcs}04$ 
pixel$^{-1}$. In 2011 the $F336W$, $F438W$, $F606W$, $F625W$, and
$F814W$ bands were employed, while in 2013 the imaging was done in the
$F275W$, $F336W$, $F438W$, $F555W$, and $F814W$ bands. Along with a
higher resolution, the new images are deeper than the pre-explosion ones. 

Processed {\em HST} data were downloaded from the Hubble Legacy
Archive\footnote{\url{http://hla.stsci.edu}} (HLA). The WFPC2 observations
were performed in pairs of exposures with each filter. The combined
$F606W$ image had poor quality due to a misalignment of both
exposures. We thus performed the image combination for that band using
the DRIZZLEPAC package. The WFC3 images from 2011 and 2013 consisted
on four and three integrations per band, respectively. 

\citet{Crockett08} obtained ground-based, high-resolution imaging of
the SN while it was still bright in order to accurately determine the
location of the SN. These observations were carried out in the $K$
band using adaptive optics with the Altair/NIRI instrument mounted on
the Gemini North Telescope. We obtained the raw images from the Gemini
Science
Archive\footnote{\url{www.cadc-ccda.hia-iha.nrc-cnrc.gc.ca/en/gsa}}
and processed them through sky subtraction and 
combination using the dedicated tools for NIRI in the {\tt gemini}
package of IRAF. The co-added image has excellent
quality, with FHWM of $0{\farcs}09$ ($\approx 4.1$ pixels).

\begin{deluxetable*}{lccccl}  
\tabletypesize{\small} 
\tablecolumns{6} 
\tablewidth{0pt} 
\tablecaption{Summary of {\em HST} observations.\label{tab:obs}} 
\tablehead{ 
\colhead{Date} & \colhead{Instrument/} & \colhead{Filter}  & \colhead {Exposure} & \colhead{Program} & \colhead{Program}\\
\colhead{} & \colhead{Detector} & \colhead{} & \colhead{(s)} & \colhead{ID} & \colhead{PI} 
}
\startdata 
\multicolumn{5}{c}{Pre-explosion} \\
1994-12-03 & WPFC2/WFC & $F606W$ & 160 & GO-5446 & Illingworth\\
2001-07-02 & WFPC2/PC & $F300W$ & 600 & GO-9124 & Windhorst\\
2001-11-13 & WFPC2/WFC & $F450W$ & 460 & GO-9042 & Smartt\\
2001-11-13 & WFPC2/WFC & $F814W$ & 460 & GO-9042 & Smartt\\
\multicolumn{5}{c}{Post-explosion} \\
2011-07-12 & WFC3/UVIS & $F336W$ & 1704 & GO-12262 & Maund\\
2011-07-12 & WFC3/UVIS & $F438W$ & 1684 & GO-12262 & Maund\\
2011-07-12 & WFC3/UVIS & $F606W$ & 1310 & GO-12262 & Maund\\
2011-07-12 & WFC3/UVIS & $F625W$ & 1029 & GO-12262 & Maund\\
2011-07-12 & WFC3/UVIS & $F814W$ & 1939 & GO-12262 & Maund\\
2013-10-29 & WFC3/UVIS & $F275W$ & 2361 & GO-13364 & Calzetti\\
2013-10-29 & WFC3/UVIS & $F336W$ & 1107 & GO-13364 & Calzetti\\
2013-10-29 & WFC3/UVIS & $F438W$ &  953 & GO-13364 & Calzetti\\
2013-10-29 & WFC3/UVIS & $F555W$ & 1131 & GO-13364 & Calzetti\\
2013-10-29 & WFC3/UVIS & $F814W$ &  977 & GO-13364 & Calzetti
\enddata 
\end{deluxetable*} 

\subsection{Distance and Extinction}
\label{sec:dist}

\noindent Distance and extinction are usually the largest sources of
uncertainty in the study of SN progenitors from direct detections. The
distance to NGC~4490 has been computed as $9.6\pm1.3$ Mpc by
\citet{Pastorello08} from an average of several measurements. This
value has been extensively used in the literature for
SN~2008ax. We note that the average includes distances derived
from the recession velocity of the host galaxy. These are relatively
large distances that involve large uncertainties given the small
recession velocity of NGC~4490. In this work we instead
adopt an updated distance of $7.77\pm1.54$ Mpc, as provided
by the NASA/IPAC Extragalactic Database (NED), based on seven
measurements done with the Tully-Fisher
\citep{Tully88,Theureau07,Karachentsev13}, and 
Sosies \citep{Terry02} methods.

Dust extinction in the direction toward SN~2008ax appears to be
significant. The Milky-Way contribution is small, the NED provides 
$E(B-V)_{\mathrm{MW}}=0.019$ mag based on infrared dust maps of
\citet{Schlafly11}. For the host-galaxy reddening, several estimates
have been given in the literature based on equivalent widths (EW) of
interstellar lines. \citet{Pastorello08} inferred a reddening of 
$E(B-V)_{\mathrm{host}}=0.3$ mag from the EW of the \ion{Na}{1}~D line
and using the empirical relation of \citet{Turatto03}. 
\citet{Chornock11} noted that the \ion{Na}{1}~D line is saturated and
used the EW of the \ion{K}{1}~$\lambda$7699 instead to derive
$E(B-V)_{\mathrm{host}}=0.5$ mag based on the relation of
\citet{Munari97}. \citet{Chornock11} also provided an EW of the diffuse
interstellar band (DIB) at 5780 \AA\ observed in high-resolution
spectra. With their value of 228 m\AA\ and the relation presented by
\citet{Phillips13}, we obtain $E(B-V)_{\mathrm{host}}=0.4\pm0.2$ mag,
assuming a standard reddening law of \citet{Cardelli89} with
$R_V=3.1$. 

A more accurate measure of the host-galaxy reddening can be obtained
from the color evolution. We compared the MW-corrected $(B-V)$ colors
of SN~2008ax with those of a sample of reddening-free
stripped-envelope SNe observed 
by the Carnegie Supernova Project \citep[CSP;][]{Hamuy06}. The
reddening-free sample is composed of SNe with no narrow \ion{Na}{1}~D
absorptions and located relatively apart of star-forming regions. An
average intrinsic $(B-V)$ color curve was constructed from this sample
between maximum light and 20 days after. The intrinsic colors evolve
monotonically during this interval from $(B-V)=0.35$ mag to $1.08$
mag, with a dispersion of $0.06$--$0.14$ mag. The color curve of
SN~2008ax reproduced the same shape but shifted toward redder
colors. We adopt the average shift as the color excess,
$E(B-V)_{\mathrm{host}}=0.27\pm0.02$ mag. Due to the dispersion in the
intrinsic colors, we consider a systematic uncertainty of $\approx0.1$
mag. 

We also compared $(V-R)$ colors evaluated at 10 days past $V$- and
$R$-band maximum light with the intrinsic-color calibrations of
\citet{Drout11} of $\langle(V-R)\rangle_{V10}=0.26 \pm 0.06$ mag and
$\langle(V-R)\rangle_{R10}=0.29 \pm 0.08$ mag, respectively. For
SN~2008ax we measure $(V-R)_{V10}=0.51\pm0.03$ mag and
$(V-R)_{R10}=0.57\pm0.03$ mag after correcting for MW reddening. An 
average of the differences from the reference values yields
$E(V-R)_{\mathrm{Host}}=0.26 \pm 0.07$ mag. Assuming a standard
extinction law with $R_V=3.1$, we obtain 
$E(B-V)_{\mathrm{Host}}=0.38 \pm 0.08$ mag. 

Considering the color excesses derived from observed colors, we adopt
a host-galaxy reddening of $E(B-V)_{\mathrm{Host}}=0.3 \pm 0.1$
mag. The same value was adopted by \citet{Crockett08} in their
analysis of the progenitor candidate, although their distance was
$\approx20$\% longer.

\section{ANALYSIS}
\label{sec:ana}

\subsection{Image Registration and Photometry}
\label{sec:reg}

\begin{figure*}[htpb] 
\includegraphics[width=0.48\textwidth]{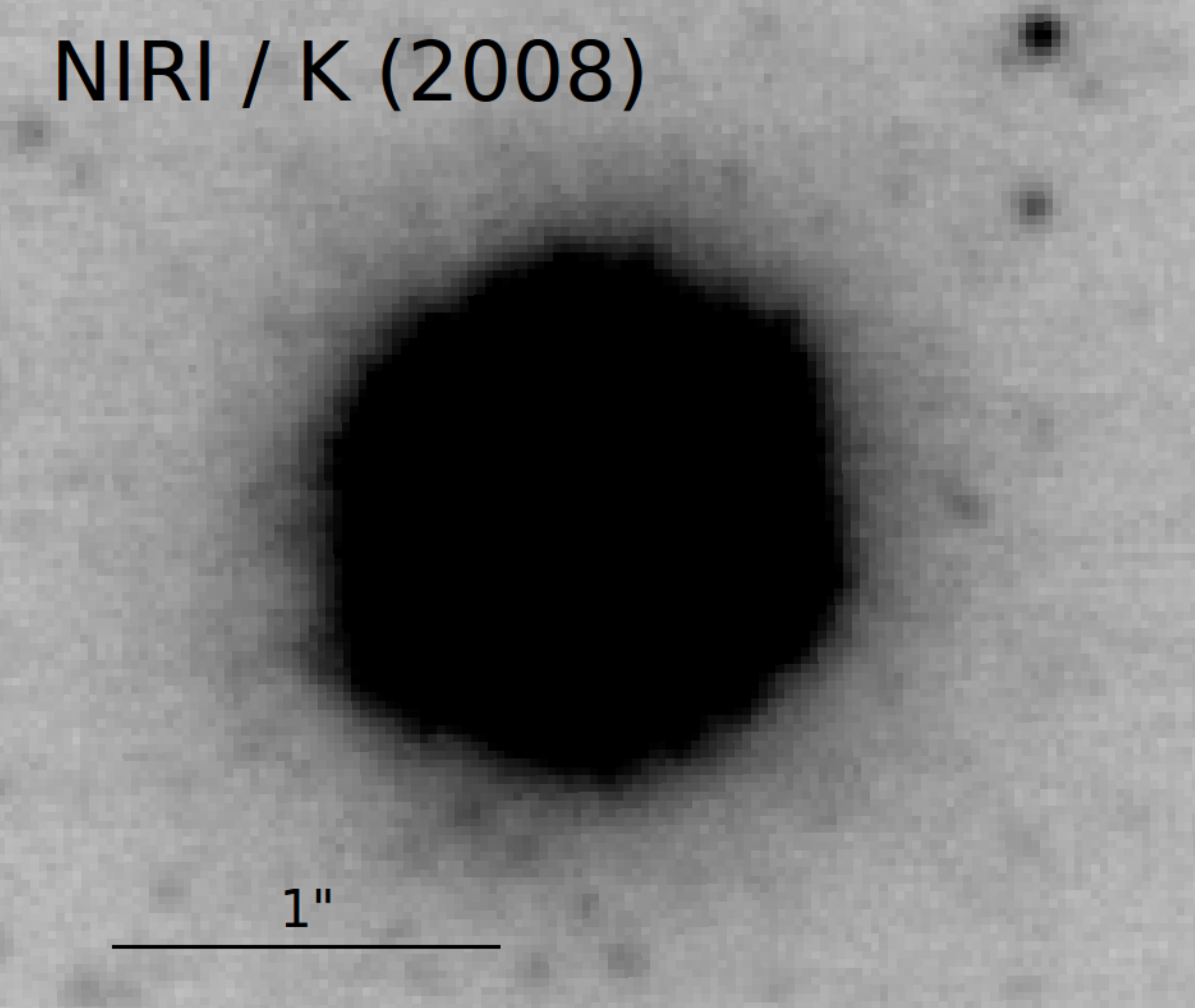}\hspace{0.03\textwidth}\includegraphics[width=0.48\textwidth]{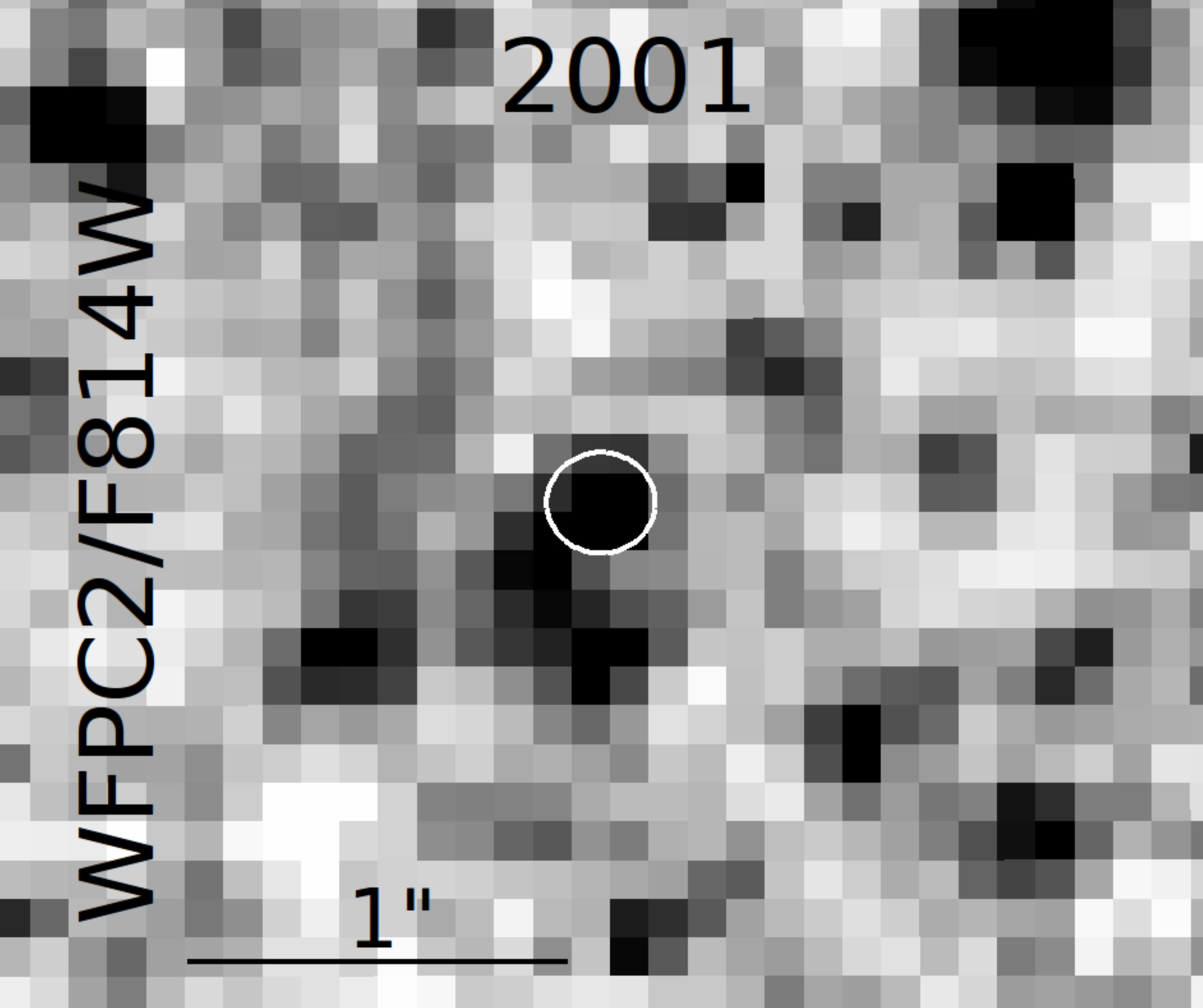}

\vspace{0.02\textheight}

\includegraphics[width=0.48\textwidth]{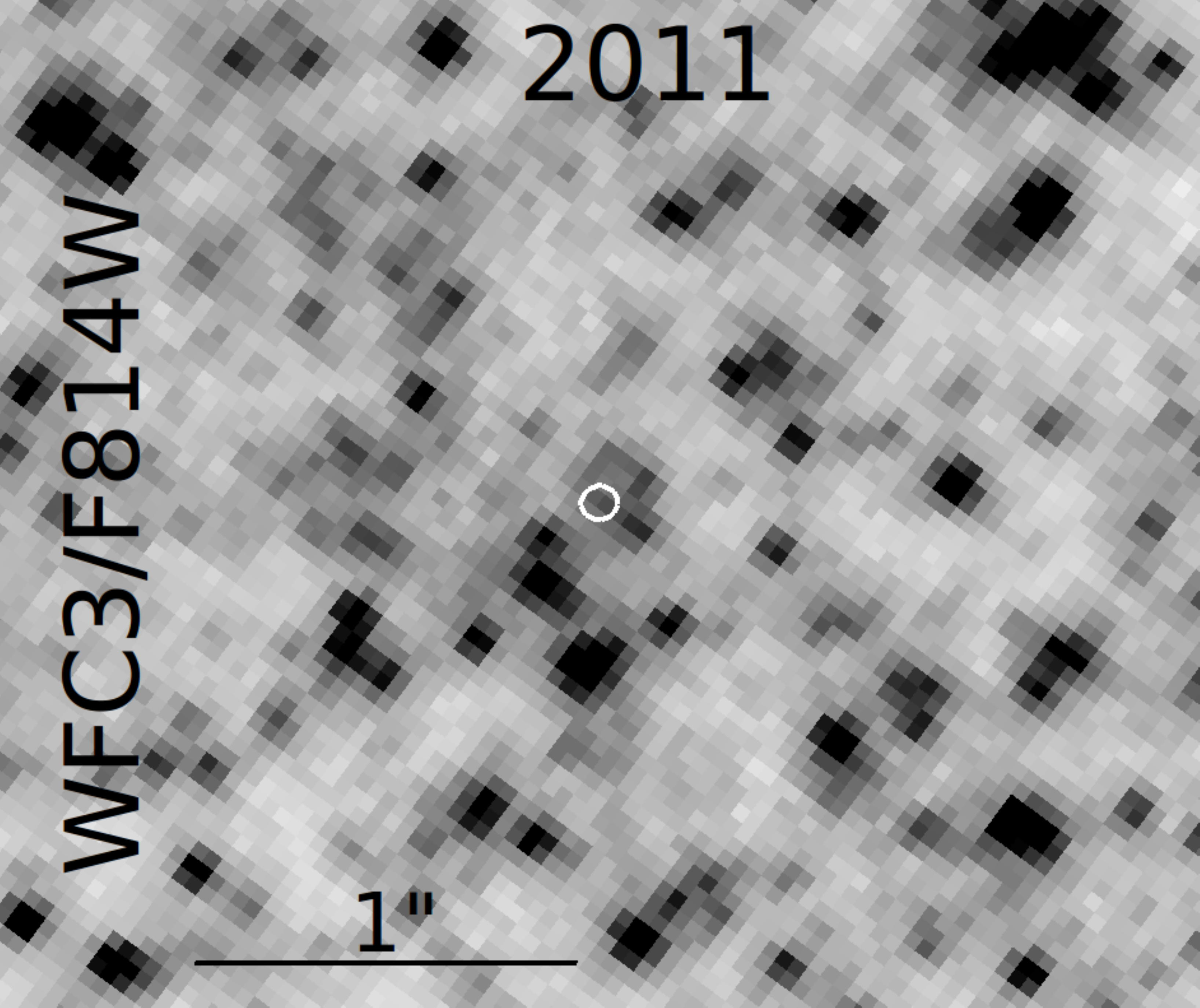}\hspace{0.03\textwidth}\includegraphics[width=0.48\textwidth]{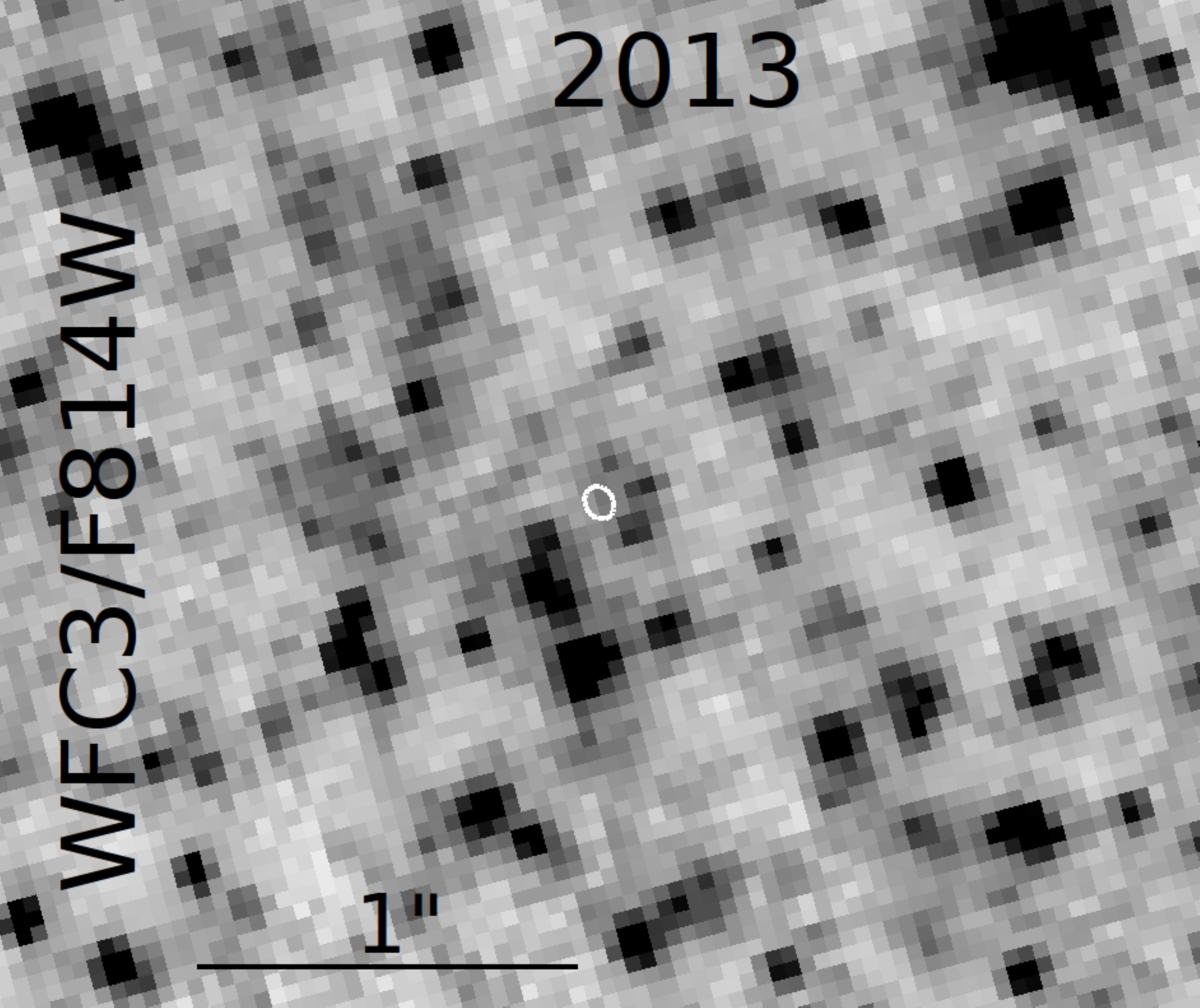}
\caption{({\em Top left}) Ground-based $K$-band image obtained in
  2008, while the SN was bright, with Altair/NIRI at Gemini North
  Telescope. ({\em Top right}) {\em HST} image of the SN site obtained
  before the explosion with WFPC2 and filter $F814W$. ({\em Bottom
    row}) {\em HST} images of the SN site obtained after the explosion
  using the WFC3/UVIS and filter $F814W$ in 2011 ({\em left}) and 2013
  ({\em right}). The four images
  are shown with the same scale of about $3{\farcs}1$ by $2{\farcs}6$
  and aligned with North up and East to the left. The SN location as
  determined from the registration with the Gemini Altair/NIRI image
  is indicated in each {\em HST} frame with a white ellipse of
  5$\sigma$ semi-axis in $x$ and $y$. The object detected in the
  pre-explosion image is resolved into four objects in 2011, one of
  which is the progenitor candidate that fades away in 2013.
  \label{fig:reg}}
\end{figure*}

\noindent We used the
combined NIRI image to find the location of the 
SN on the $F814W$ images obtained at different epochs. We employed 39
stars in common---the same stars at all epochs---to align the {\em
  HST} images with respect to the NIRI image. The registration
produced $x,y$ rms uncertainties of $28,26$ mas ($0.28,0.26$ pixel),
$8.8,9.6$ mas ($0.22,0.24$ pixel), and $7.2,11.2$ mas ($0.18,0.28$
pixel) for the 2001, 2011, and 2013 images, respectively. The location
of the SN on the $F814W$ images is shown in Figure~\ref{fig:reg} and
corresponds to pixels $[208.07,553.01]$, $[512.57,556.96]$, and
$[3516.25,2246.40]$ in the WFPC2 image, the 2011 WCF3 image and the
2013 WFC3 image, respectively. We confirm the identification of the
pre-SN object detected by \citet{Crockett08} in the WFPC2 images. But
the late-time WFC3 images reveal critical information previously
unavailable about the possible SN progenitor. 

The first thing that strikes the eye from Figure~\ref{fig:reg} is that the
object detected by \citet{Crockett08} is in fact resolved into four
sources in the 2011 image. The source to the SE matches the location
of the SN to within $\approx$1$\sigma$. We identify this source with the
fading SN. Indeed, the 2013 images show that this source has
disappeared within the noise. We indicate the other three sources with
letters A, B, and C. The object identified by
\citet{Crockett08} as the progenitor candidate is thus contaminated by
the light from objects A, B, and C. 

\begin{figure*}[htpb] 
\begin{center}
{\bf \large \hspace{-0.05\textwidth}pre-explosion\hspace{0.23\textwidth}2011\hspace{0.25\textwidth}2013}\\
\hspace{0.31\textwidth}\includegraphics[width=0.26\textwidth]{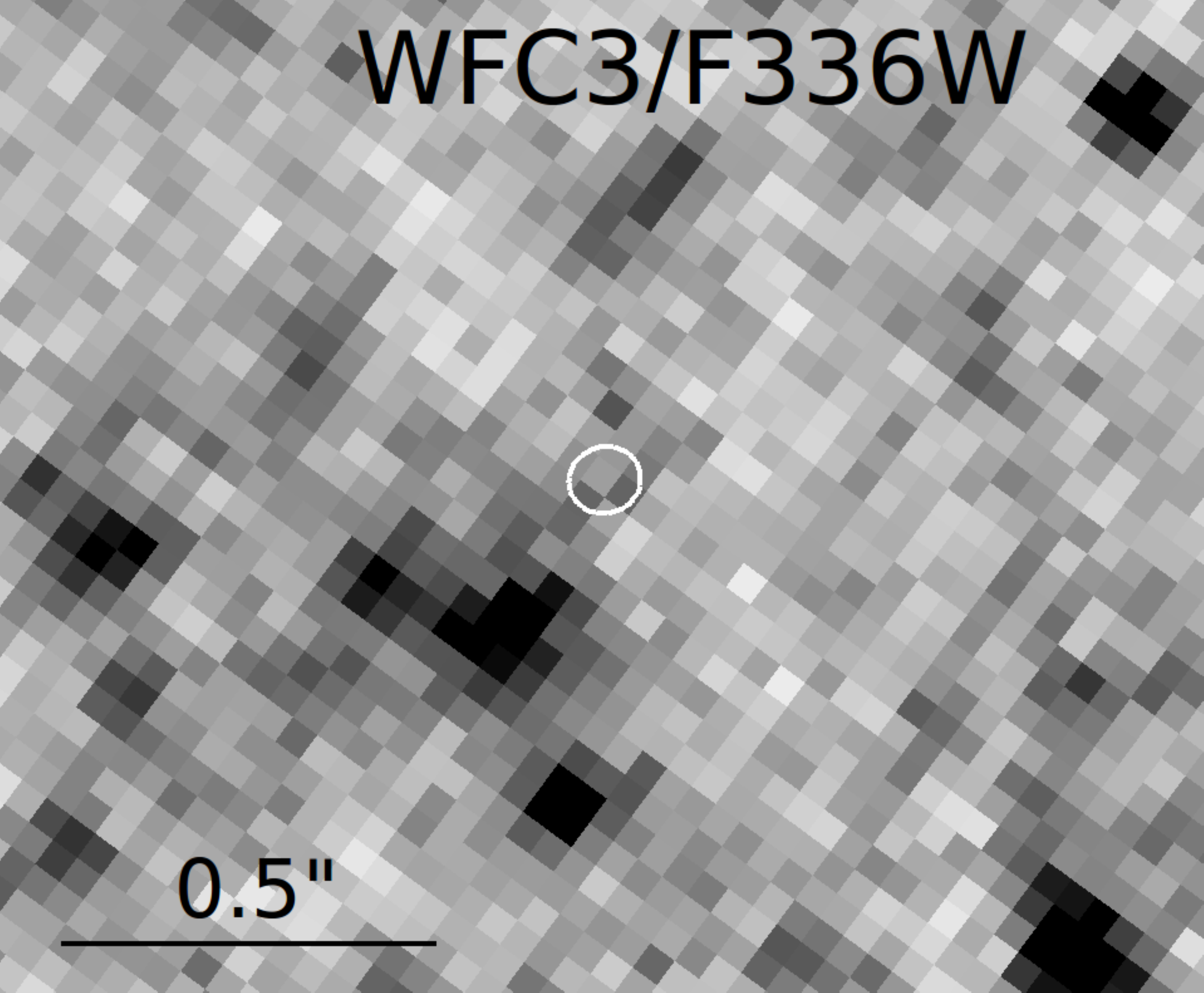}\hspace{0.05\textwidth}\includegraphics[width=0.26\textwidth]{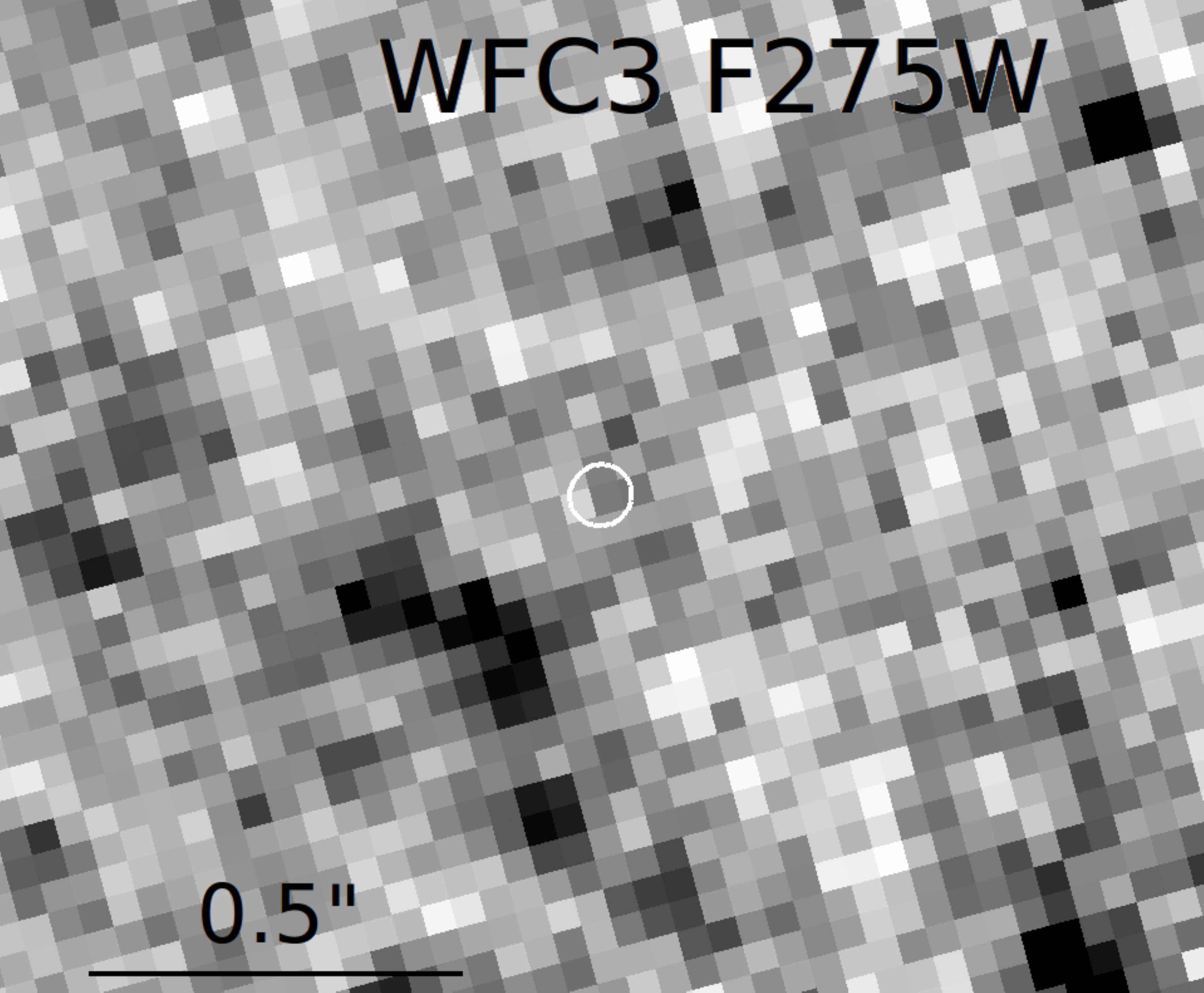}

\vspace{0.01\textheight}

\includegraphics[width=0.26\textwidth]{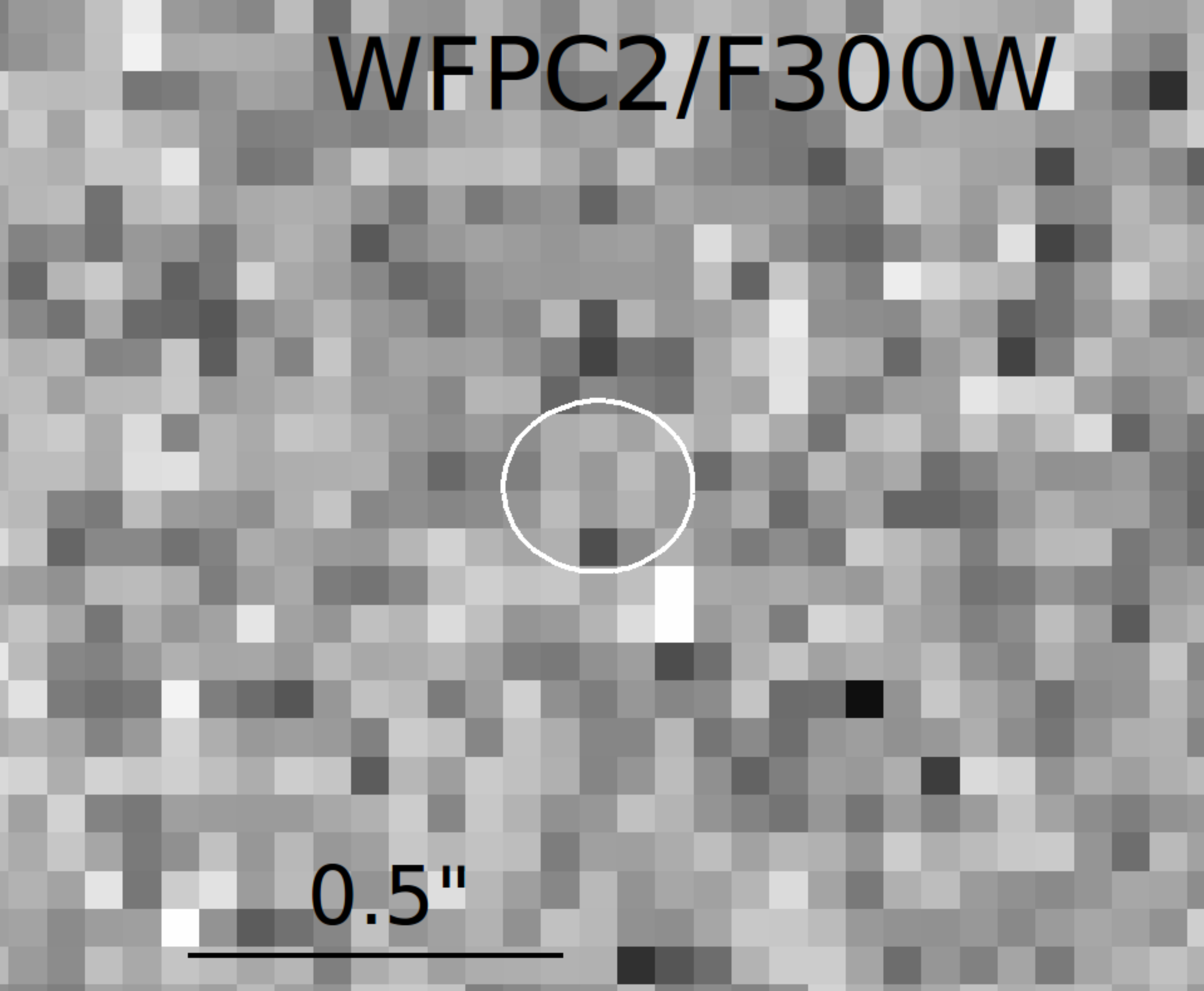}\hspace{0.05\textwidth}\includegraphics[width=0.26\textwidth]{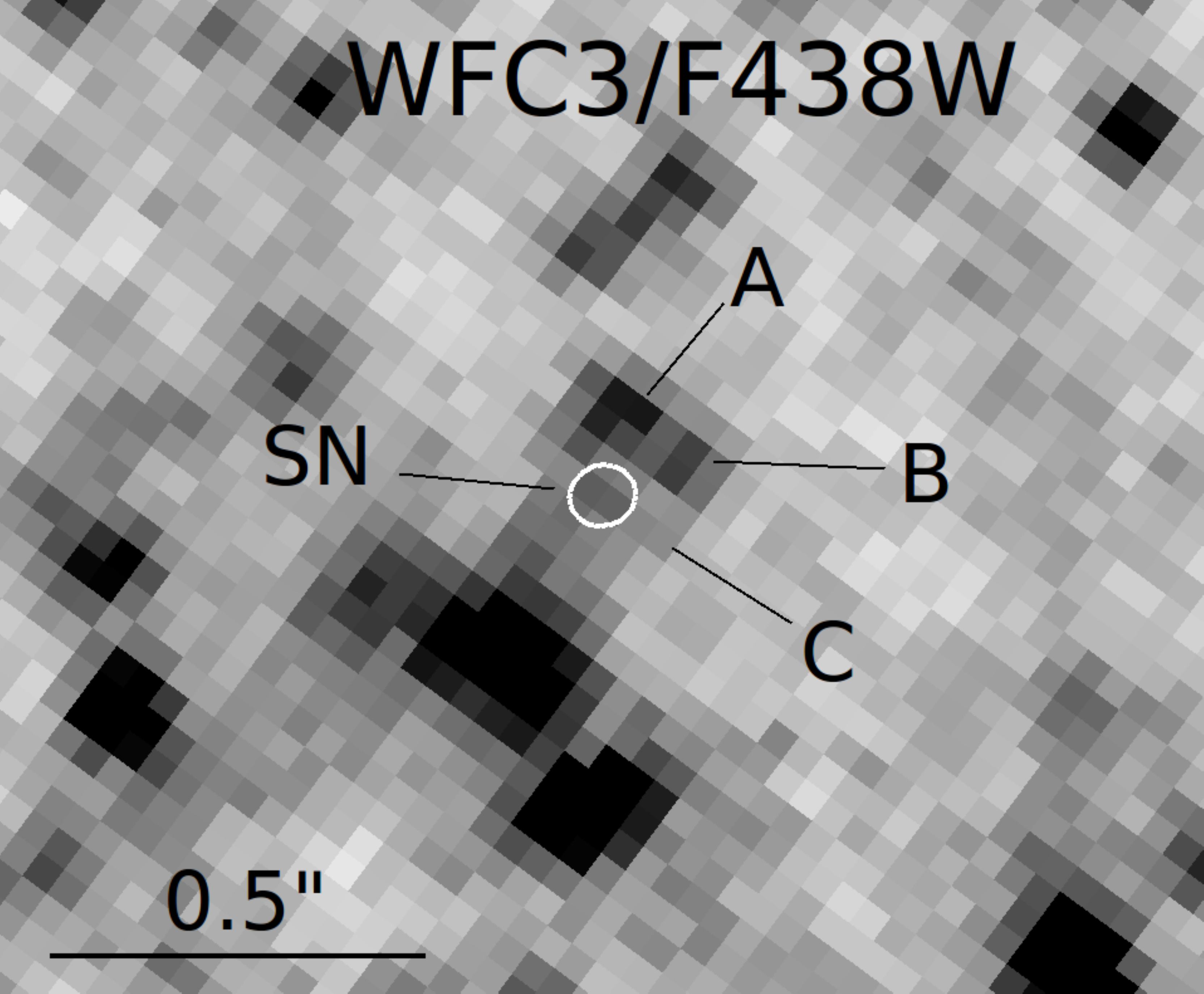}\hspace{0.05\textwidth}\includegraphics[width=0.26\textwidth]{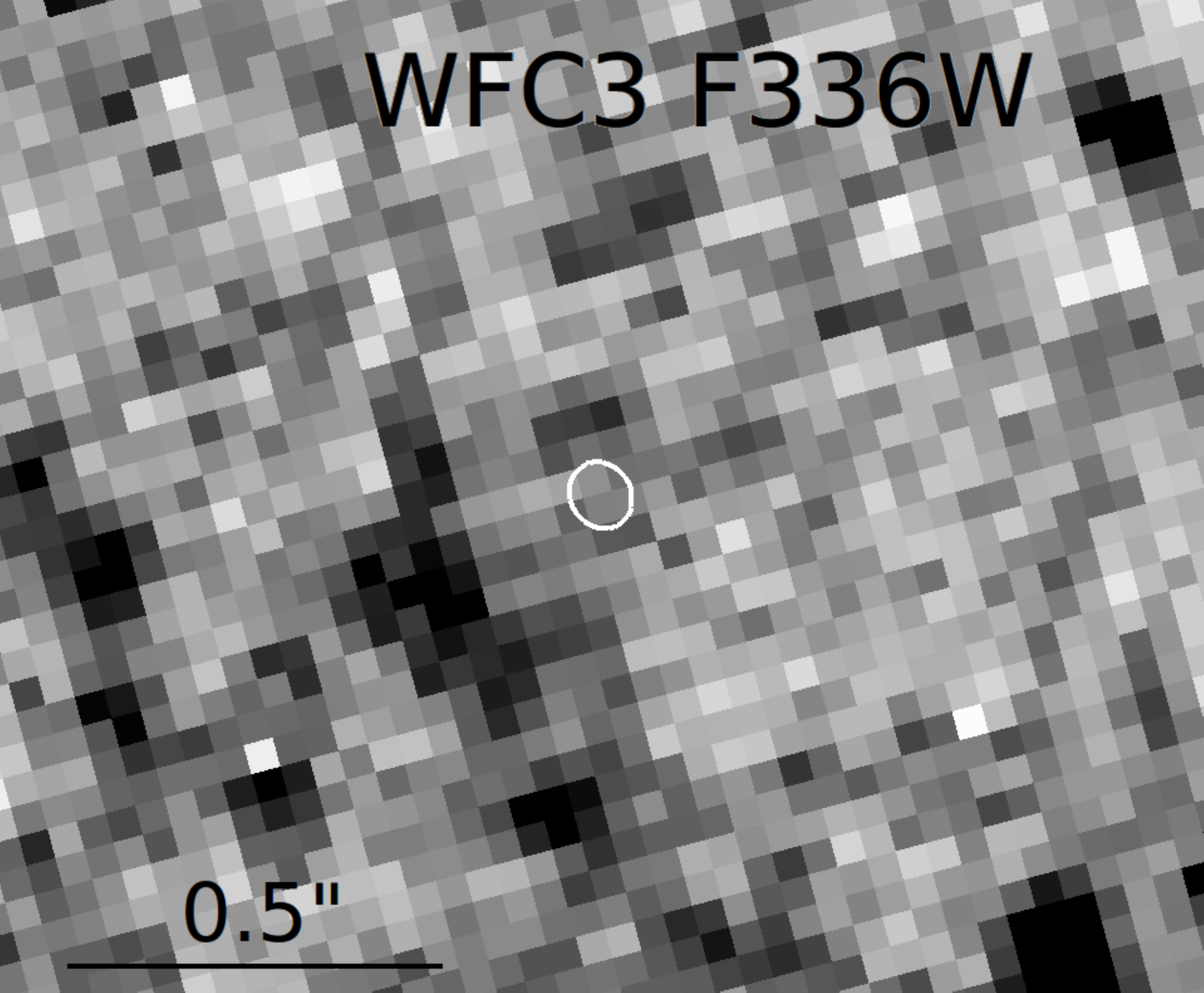}

\vspace{0.01\textheight}

\includegraphics[width=0.26\textwidth]{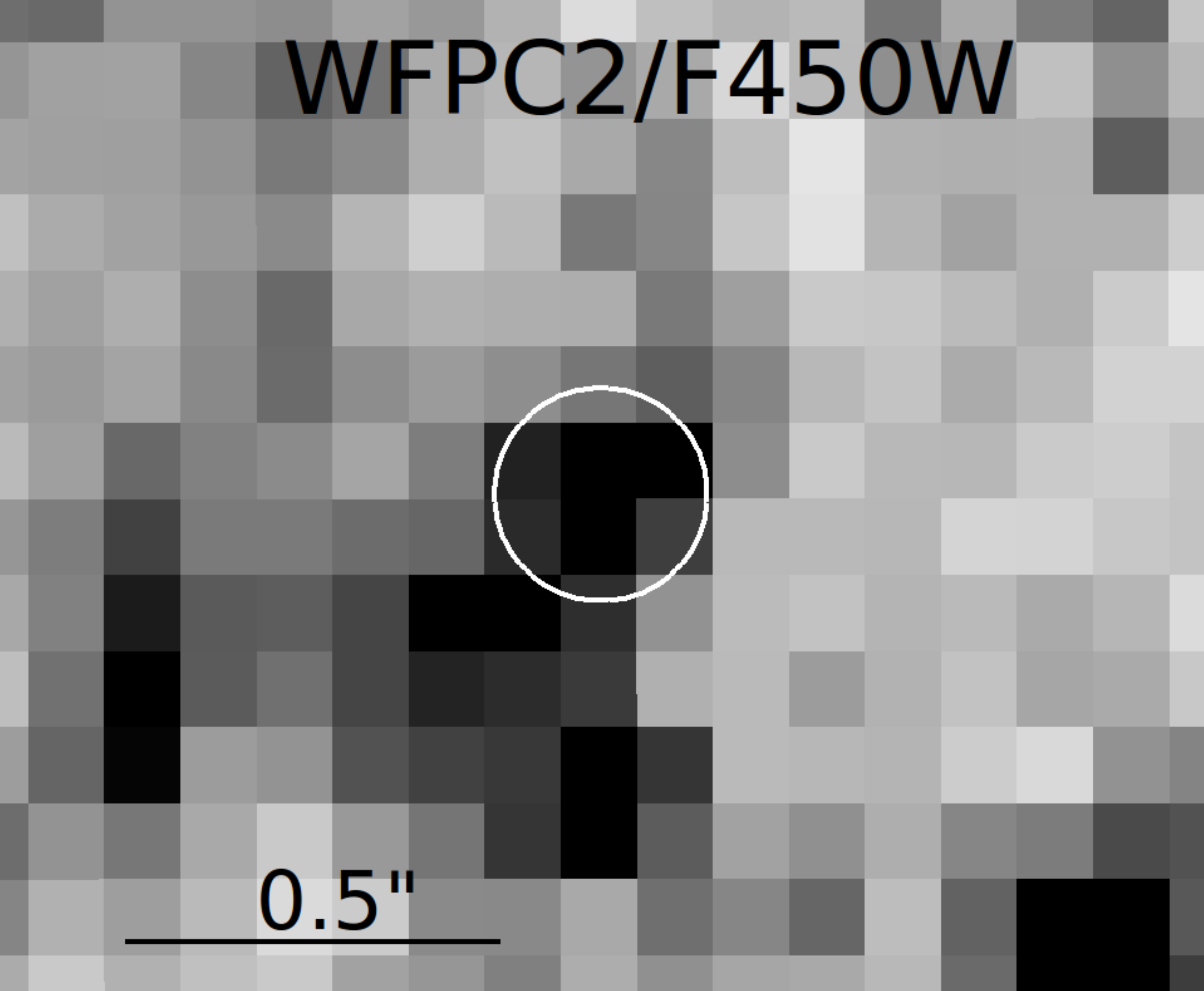}\hspace{0.05\textwidth}\includegraphics[width=0.26\textwidth]{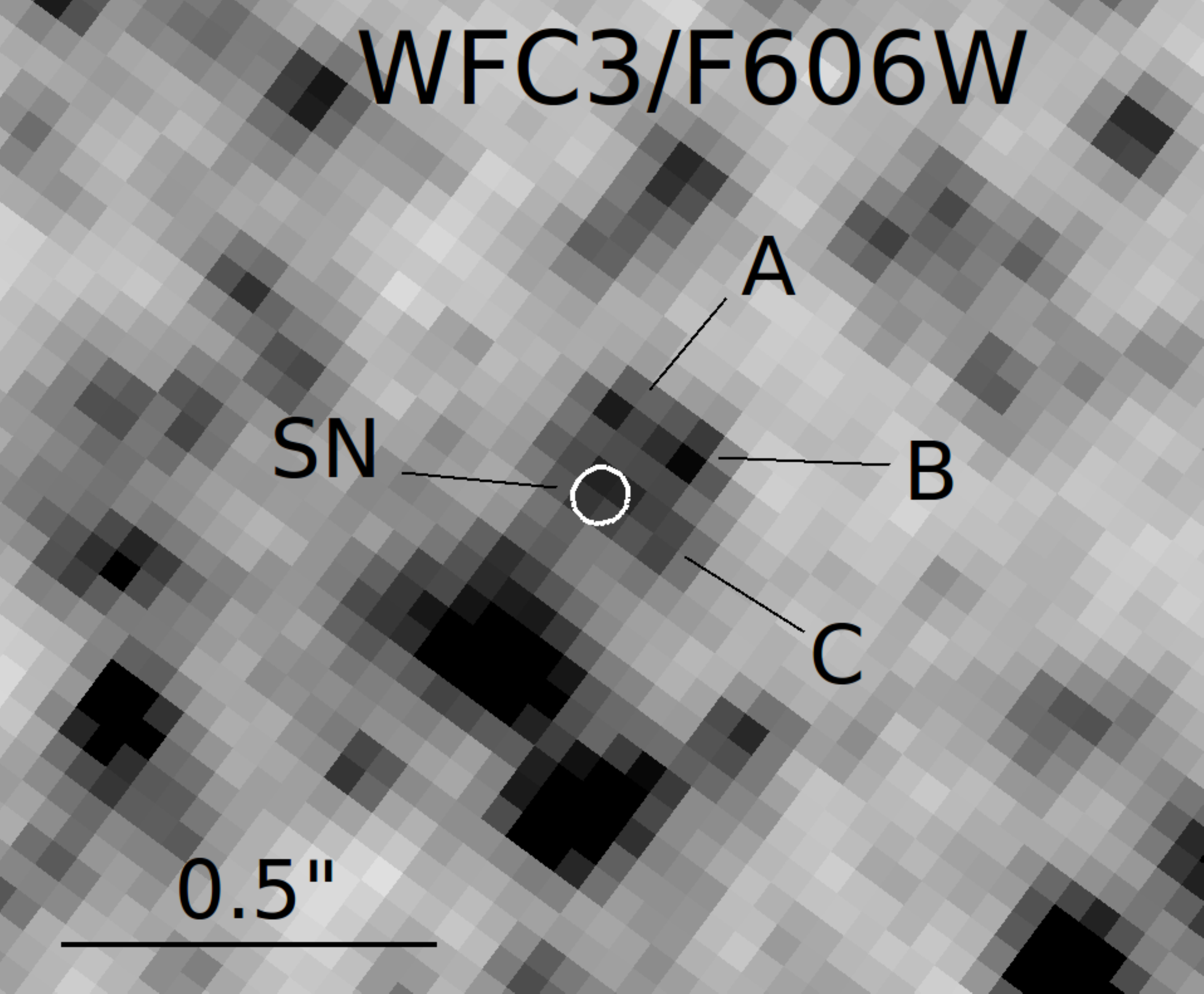}\hspace{0.05\textwidth}\includegraphics[width=0.26\textwidth]{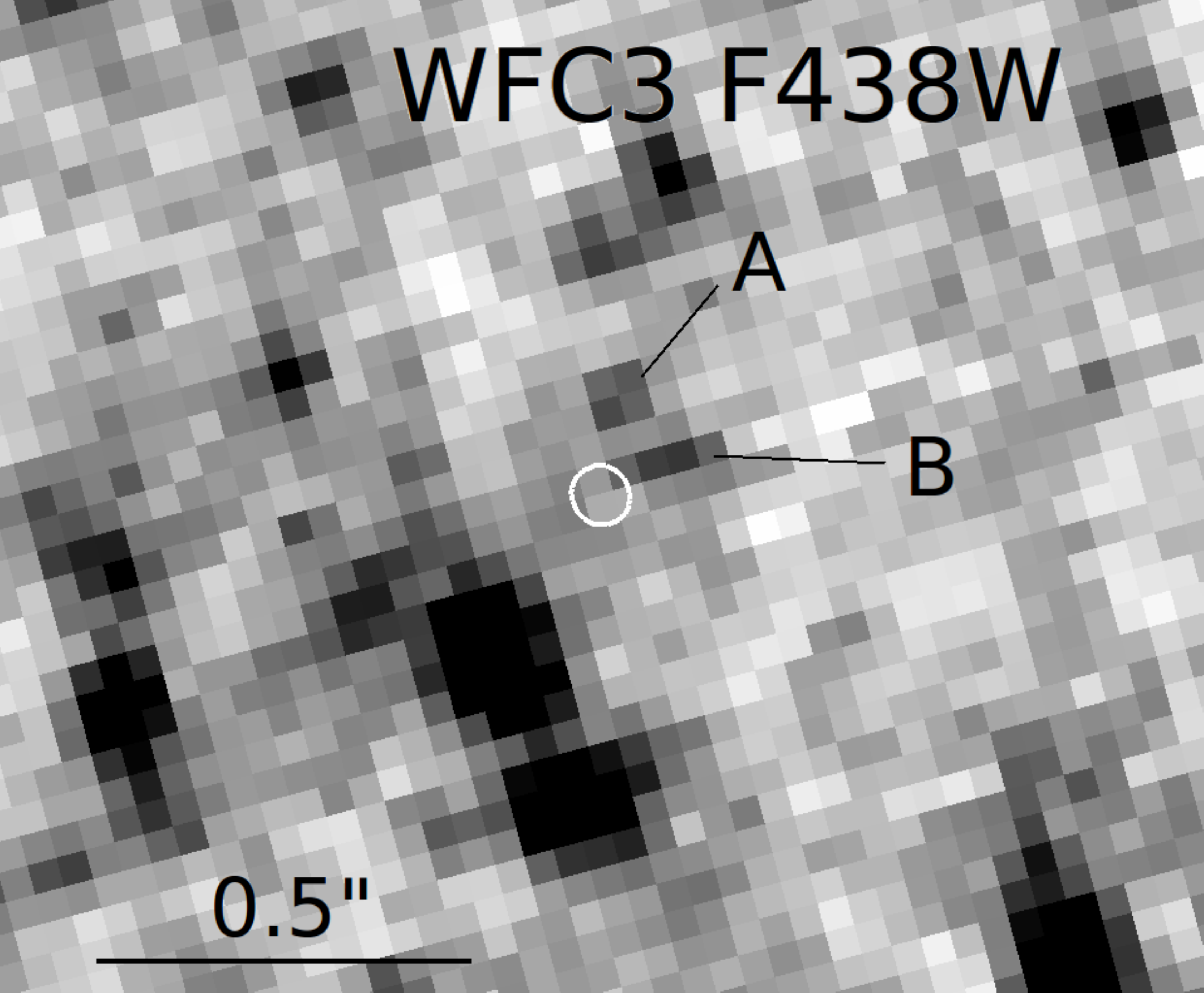}

\vspace{0.01\textheight}

\includegraphics[width=0.26\textwidth]{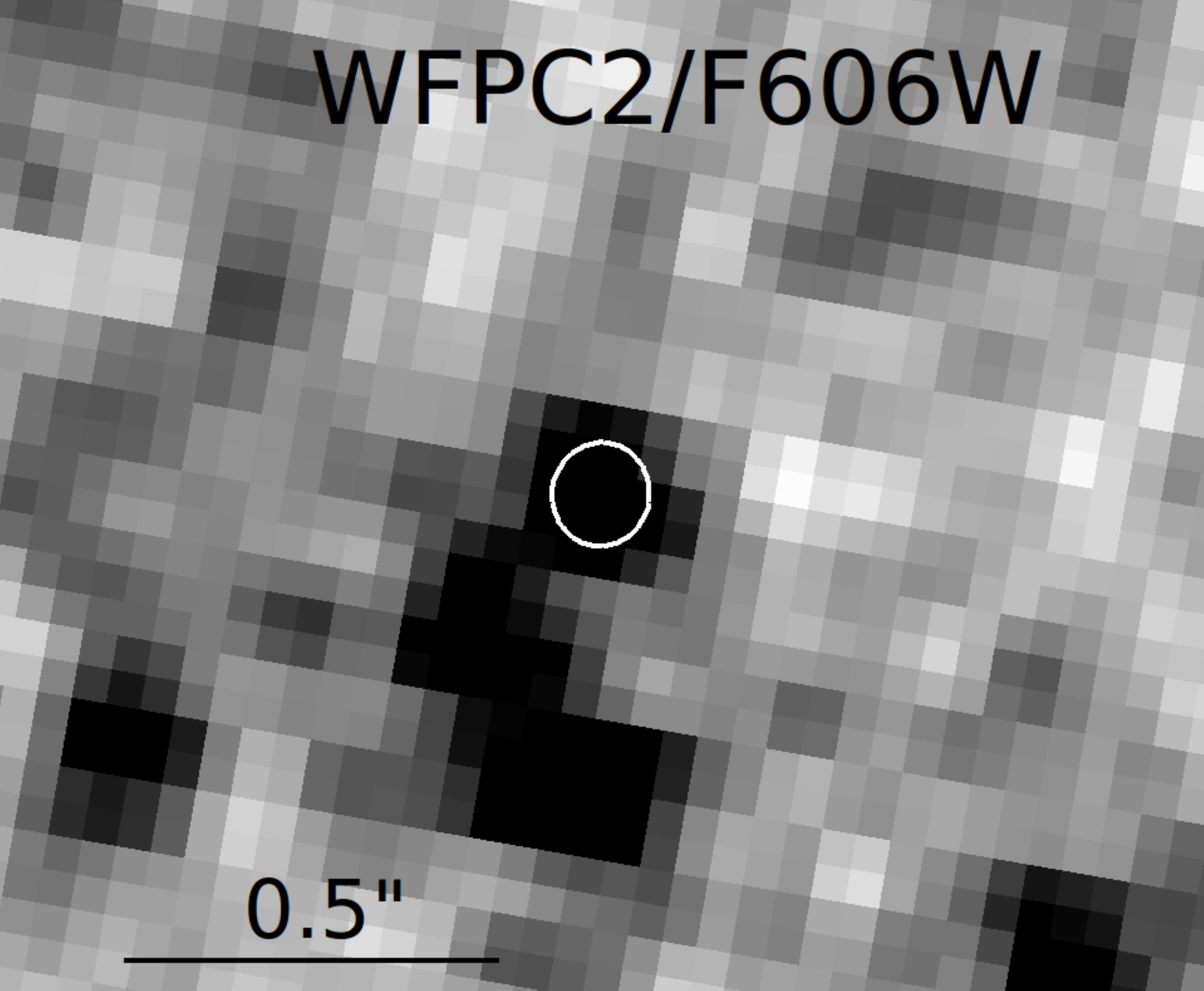}\hspace{0.05\textwidth}\includegraphics[width=0.26\textwidth]{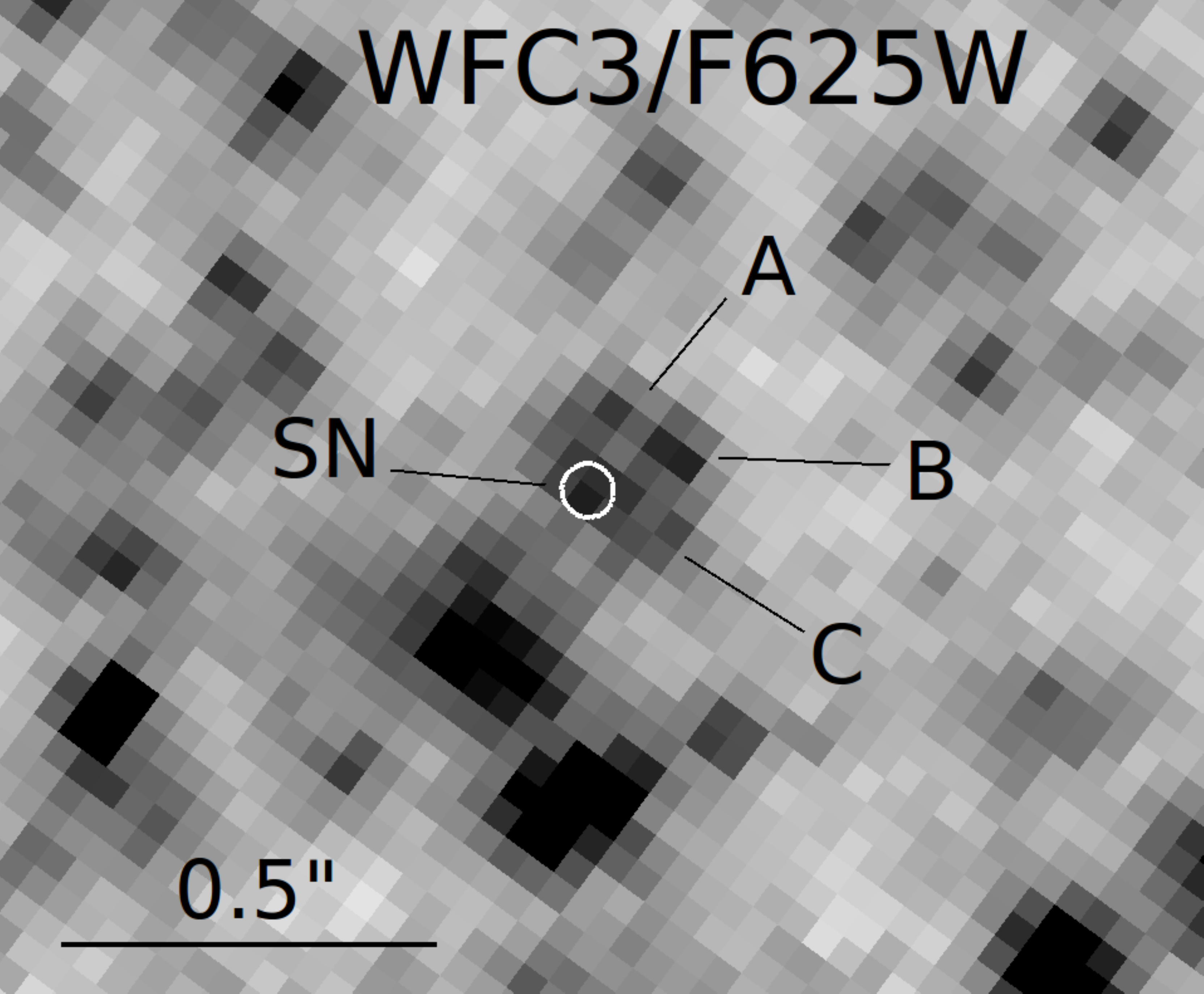}\hspace{0.05\textwidth}\includegraphics[width=0.26\textwidth]{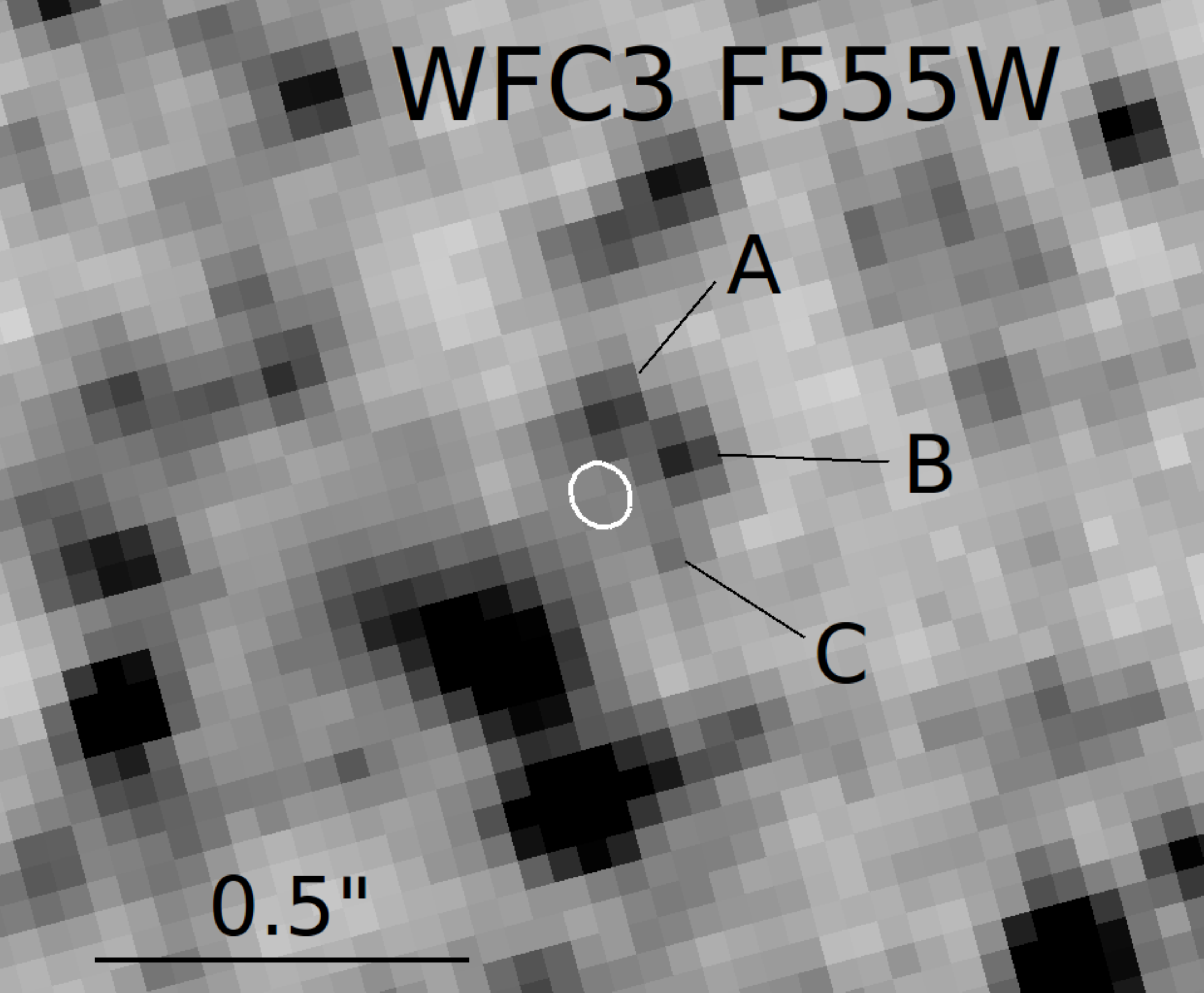}

\vspace{0.01\textheight}

\includegraphics[width=0.26\textwidth]{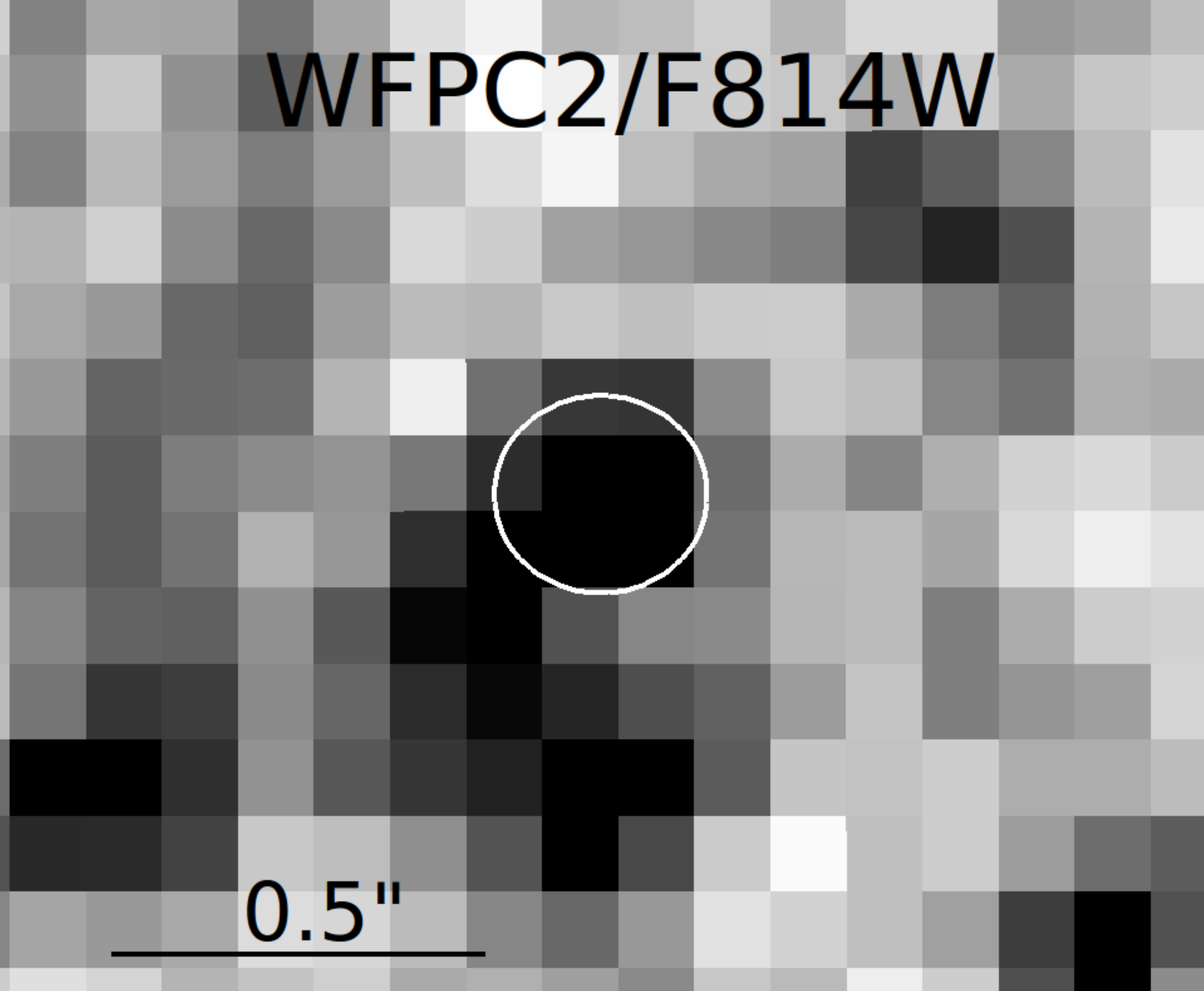}\hspace{0.05\textwidth}\includegraphics[width=0.26\textwidth]{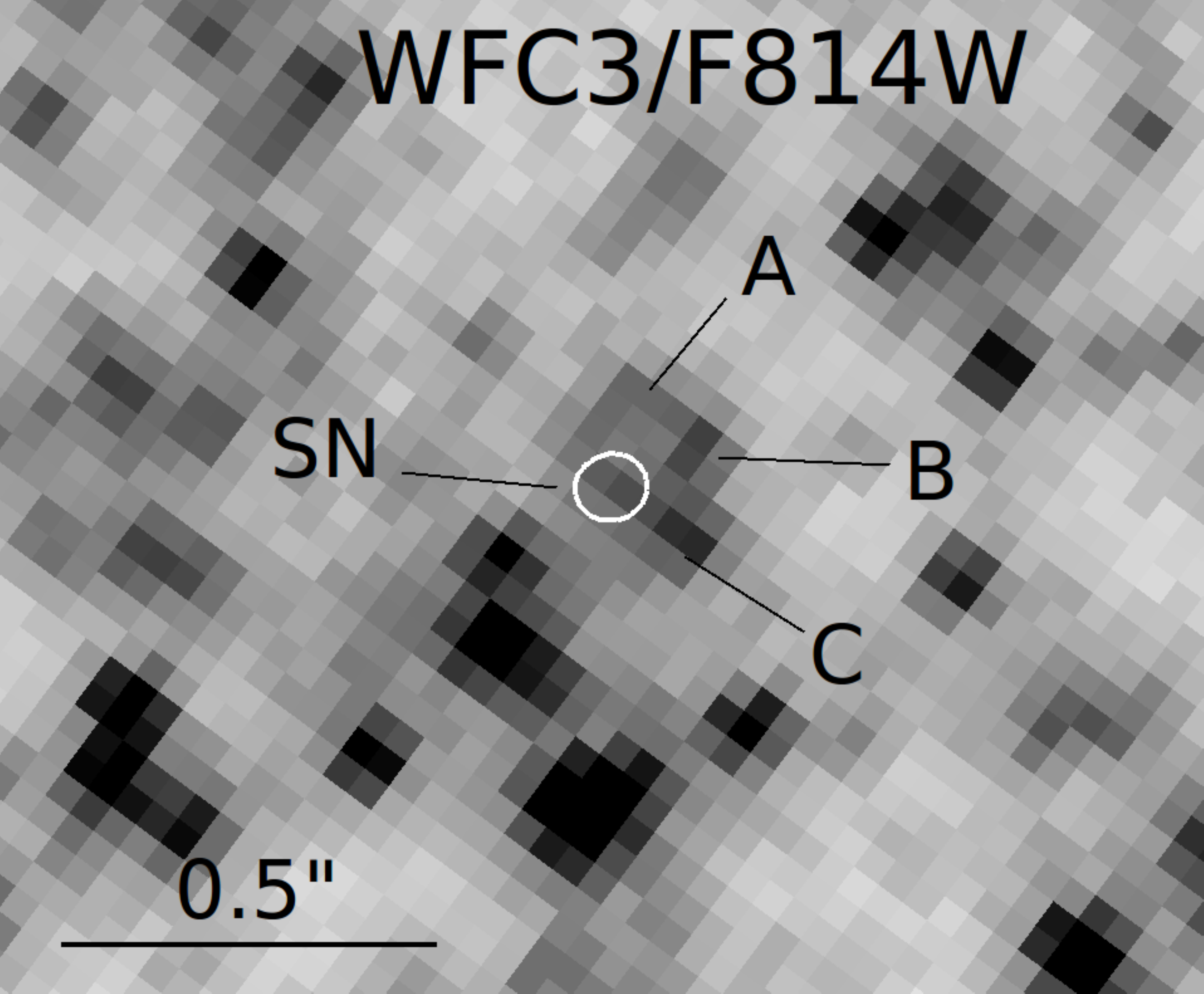}\hspace{0.05\textwidth}\includegraphics[width=0.26\textwidth]{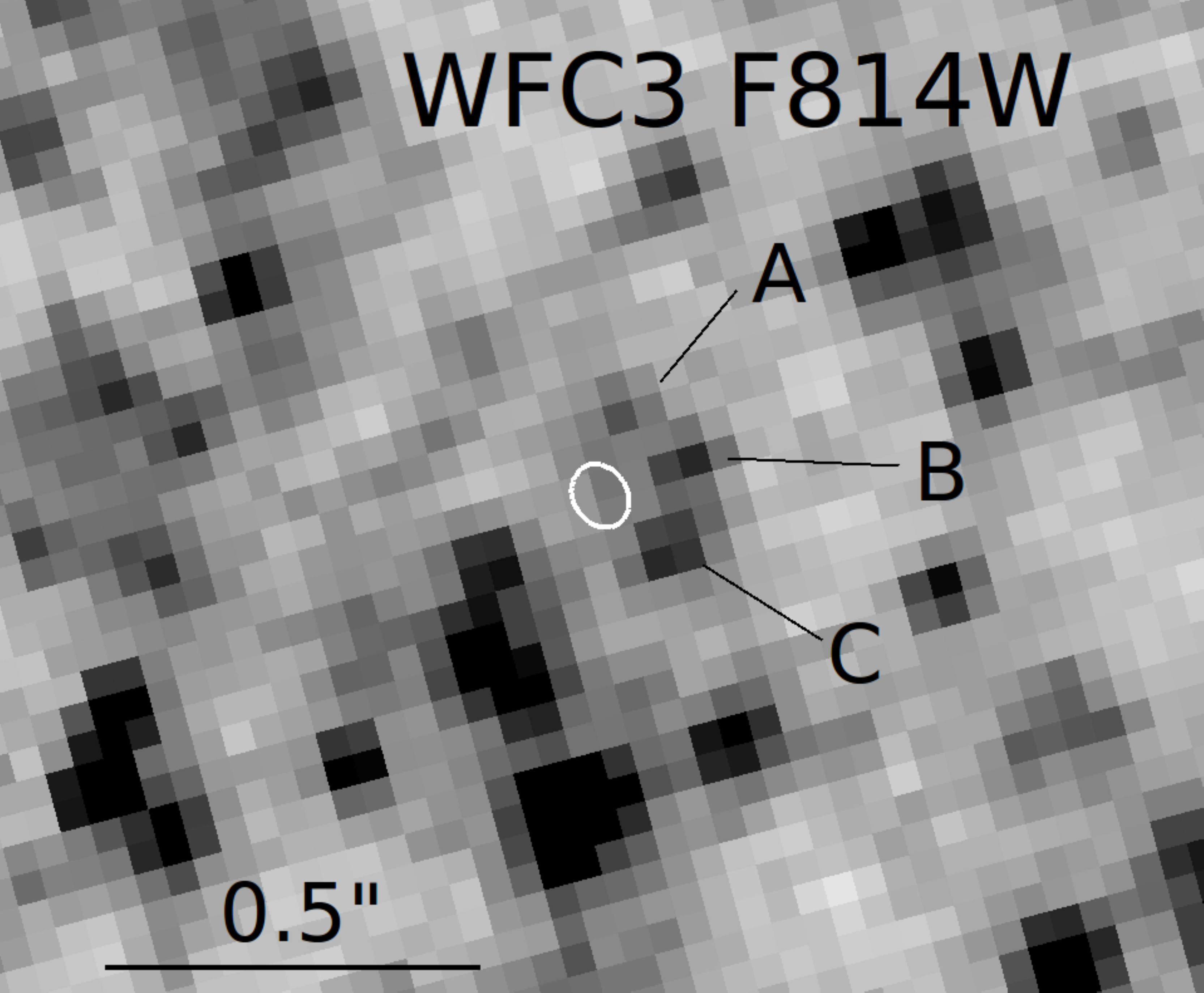}
\caption{{\em HST} images of the field of SN~2008ax. All images
  are oriented with North up and East to the left, and the field size
  is approximately $1{\farcs}6$ by $1{\farcs}3$. ({\em Left column})
  Pre-explosion images obtained in 
    1994/2001 with WFPC2. ({\em Middle column}) Post-explosion images
    obtained in 2011 with WFC3/UVIS. ({\em Right 
      column}) Post-explosion images obtained in 2013 with
    WFC3/UVIS. All images have been registered to the Altair/NIRI
    image and the location of the SN is indicated with a 5$\sigma$
    error ellipse. The SN is still detected in 2011 but it is below the
    detection limit in the shallower images from 2013. Three stars
    have been identified in the 2011 and 2013 WFC3 
    images that appear blended to the SN progenitor in the
    coarser-resolution WFPC2 pre-explosion images. Their locations in
    the images where they could be detected are indicated with letters
    ``A'', ``B'', and ``C''. 
  \label{fig:hstim}}
\end{center}
\end{figure*}

Figure~\ref{fig:hstim} shows the complete set of images used in this work.
The identification of sources A, B, and C can be done in the rest of
the WFC3 images from 2011 and 2013, except for the bluest bands where
the signal-to-noise ratio ($SNR$) is too low. We performed point-spread
function (PSF) photometry of the pre-SN source present in the WFPC2
images, and of the four sources identified in the WFC3 images. When
the objects were not detected, we estimated limiting magnitudes as
explained below.

The pre-explosion WFPC2 images were measured with the
DAOPHOT package included in IRAF\footnote{IRAF, the Image Reduction
and Analysis Facility, is distributed by the National Optical Astronomy
Observatory, which is operated by the Association of Universities
for Research in Astronomy (AURA), Inc., under cooperative agreement
with the National Science Foundation (NSF); see
\url{http://iraf.noao.edu}.}. We employed the combined frames
except in the $F606W$ band which showed poor alignment of the
individual exposures. In that band we measured PSF photometry on the
two images and averaged the results. The resulting photometry is
listed in Table~\ref{tab:prph} (second row), compared with that obtained by
\citet{Crockett08} on the same images (first row). The values are
consistent within the uncertainties. We obtain a larger uncertainty in
the $F606W$ band, probably due to a larger inconsistency between the
two individual exposures. We thus confirm that the magnitude in this
band is about $0.3$ mag brighter than what was originally reported by
\citet{Li08} who may have used the misaligned image from the HLA. 

No object was detected at the SN location in the WFPC2/$F300W$ image
(Figure~\ref{fig:hstim}). We thus estimated a limiting magnitude 
from the pixel statistics in the region of the PC chip around the SN
site. Following \citet{Harris90}, we defined the limiting magnitude
for $SNR=5$ as that of an object with peak 
pixel value of three times the standard deviation of the background
($\sigma_{\mathrm{bkg}}=0.008$ counts s$^{-1}$). Due to the lack of
point sources in the PC chip, we could not compute a suitable PSF that
would allow us to convert the detection threshold to a 
magnitude. Instead, we used the encircled energy fractions tabulated
by \citet{Holtzman95} for apertures of different radii. 
From Table~2(a) in
\citet{Holtzman95}, the fraction of flux in a radius of 1 pixel versus
120 pixels is $f_{1/120}=0.33$ and $0.38$ at the $F255W$ and $F336W$ bands,
respectively, with a typical uncertainty of $0.05$. We assumed an
average was valid for $F300W$, and thus the limiting flux was
$F_{\mathrm{lim}}(F300W)=3\,\sigma_{\mathrm{bkg}}\,\pi\,/f_{1/120}=0.21$
counts s$^{-1}$. Using the flux calibration of 
WFPC2 \citep[Table~5.1 in][]{Gonzaga10} 
the limiting Vega-system magnitude corresponding to that flux is
$m_{\mathrm{lim}}(F300W)=-2.5\,\log(F_{\mathrm{lim}}(F300W)\times\mathrm{PHOTFLAM})
- \mathrm{Vega}_{\mathrm{ZP}}$. With 
$\mathrm{PHOTFLAM}=6.137\times 10^{-17}$ and
$\mathrm{Vega}_{\mathrm{ZP}}=19.406$ mag, this is
$m_{\mathrm{lim}}(F300W)=22.8$ mag. This value is similar to that of
$22.9$ mag obtained by \citet{Crockett08}

On the WFC3 images we used the DOLPHOT v2.0 package
\citep{Dolphin00}. DOLPHOT provides a magnitude from each exposure. We
only considered those measurements where the object was detected in
all frames. In those cases, the final magnitude was obtained as a
weighted average of the individual measurements. In other cases we
estimated a limiting magnitude based on the local pixel
statistics. The detection limit was set as three times the background standard
deviation, and this value was converted into a limiting magnitude in
the Vega system using the PSF computed for the image. 
The resulting magnitudes and limits are listed in
Table~\ref{tab:phot}. The photometry of objects A, B, and C remains
stable within the uncertainties between 2011 and 2013, while the
object identified as the SN in 2011 is no longer detected in
2013. The limits at this latter epoch indicate that the
SN became, as expected, fainter with time. 

Objects A, B, and C are separated from the
SN by $0{\farcs}105$, $0{\farcs}115$, and $0{\farcs}113$,
respectively, as measured in the F814W image. At the distance of
NGC~4490, these correspond to projected separations of $3.9\pm0.9$ pc, 
$4.3\pm0.9$ pc, and $4.3\pm0.9$ pc. Their
profiles are compatible with being point sources, although the image
resolution with full width at half-maximum of $\mathrm{FWHM} \approx
0{\farcs}08$ corresponds to a projected diameter of $\approx$3 pc. Their
relative brightness and separation from the SN site make 
them blend into one source in the WFPC2 images (see
Figure~\ref{fig:reg}). As already indicated by \citet{Crockett08},
the size of the source in the WFPC2 images is 
actually compatible with it holding more than one star.

\begin{deluxetable*}{lccccl}  
\tabletypesize{\small} 
\tablecolumns{6} 
\tablewidth{0pt} 
\tablecaption{Revised photometry of the pre-SN object\label{tab:prph}} 
\tablehead{ 
\colhead{Measurement} & \colhead{$F300W$} & \colhead{$F450W$} & \colhead{$F606W$} & \colhead{$F814W$} & \colhead{Reference} \\ 
\colhead{} & \colhead{(mag)} & \colhead{(mag)} & \colhead{(mag)} & \colhead{(mag)} & \colhead{}
}
\startdata 
Contaminated\tablenotemark{a}   & $>22.9$\tablenotemark{c} & $23.66(10)$ & $23.36(10)$ & $22.63(10)$ & \citet{Crockett08}\\
Contaminated\tablenotemark{a}   & $>22.8$\tablenotemark{c} & $23.69(13)$ & $23.25(24)$ & $22.73(09)$ & This work\\
Uncontaminated\tablenotemark{b} & $>22.9$\tablenotemark{c} & $24.14(22)$ & $23.85(42)$ & $23.61(22)$ & This work
\enddata 
\tablecomments{Uncertainties given in parentheses in units of
  $1/100$th of magnitude.}
\tablenotetext{a}{Photometry of the object detected in the pre-SN
  WFPC2 images.} 
\tablenotetext{b}{Revised photometry after subtraction of contributions from
  stars A, B and C.} 
\tablenotetext{c}{3$\sigma$ detection limit.}
\end{deluxetable*} 

\begin{deluxetable*}{lccccccc}  
\tabletypesize{\small} 
\tablecolumns{8} 
\tablewidth{0pt} 
\tablecaption{Photometry from post-explosion images.\label{tab:phot}} 
\tablehead{ 
\colhead{Date} & \colhead{$F275W$} & \colhead{$F336W$}  & \colhead{$F438W$} & \colhead{$F555W$} & \colhead{$F606W$} & \colhead{$F625W$} & \colhead{$F814W$} 
}
\startdata 
\multicolumn{8}{c}{SN~2008ax} \\
2011-07-12 & $\cdots$ & 25.836(291) & 26.085(125) & $\cdots$ & 25.168(056) & 24.877(110) & 24.825(069) \\
2013-10-29 & $>25.6$ & $>25.7$ & $>26.6$ & $>26.9$ & $\cdots$ & $\cdots$ & $>25.6$ \\
\multicolumn{8}{c}{Star A} \\
2011-07-12 & $\cdots$ & 26.063(326) & 25.529(073) & $\cdots$ & 25.333(123) & 25.257(133) & 24.847(111) \\
2013-10-29 & $>25.6$ & $>25.7$ & 25.984(177) & 25.621(176) & $\cdots$ & $\cdots$ & 25.065(160) \\
Average & $>25.6$ & 26.063(326) & 25.613(103) & 25.621(176) & 25.333(123) & 25.257(133) & 24.898(096) \\
\multicolumn{8}{c}{Star B} \\
2011-07-12 & $\cdots$ & $>26.2$ & 25.903(118) & $\cdots$ & 25.223(048) & 25.112(073) & 24.679(061) \\
2013-10-29 & $>25.6$ & $>25.7$ & 26.033(203) & 25.559(073) & $\cdots$ & $\cdots$ & 24.831(103) \\
Average & $>25.6$ & $>26.2$ & 25.954(083) & 25.559(073) & 25.223(048) & 25.112(073) & 24.719(053) \\
\multicolumn{8}{c}{Star C} \\
2011-07-12 & $\cdots$ & $>26.2$ & 27.468(396) & $\cdots$ & 25.619(081) & 25.244(135) & 24.348(047) \\
2013-10-29 & $>25.6$ & $>25.7$ & $>26.6$ & 26.220(131) & $\cdots$ & $\cdots$ & 24.379(077) \\
Average & $>25.6$ & $>26.2$ & 27.468(396) & 26.220(131) & 25.619(081) & 25.244(135) & 24.356(040)
\enddata 
\tablecomments{Uncertainties given in parentheses in milimagnitudes.}
\end{deluxetable*} 

\subsection{Spectral Energy Distribution}
\label{sec:sed}

\noindent In order to remove the contamination of the three
neighboring sources from the proposed SN progenitor found in the
pre-explosion WFPC2 images, we determined their spectral energy
distributions (SED). For this purpose, we transformed 
the magnitudes into specific fluxes at the effective wavelength of
each band. The results were put in absolute scale by adopting a
distance modulus of $29.42\pm0.42$ mag as given by NED. We further
assumed the extinction to be the same for the four objects as that of
derived for the SN in Section~\ref{sec:dist}. The corrected SED for 
objects A, B, and C are shown in the left panel of Figure~\ref{fig:sed}. 

The SED were fit with stellar atmosphere models from
\citet{Kurucz93}. The results are shown in Figure~\ref{fig:sed} and 
the corresponding effective temperatures and luminosities are given in
Table~\ref{tab:sed}. These fits indicate that the objects are
compatible with being single stars. The sum of the SED from stars A,
B, and C was assumed to have contaminated the photometry of the
pre-explosion object detected in the WFPC2 images. 
The total A+B+C SED flux was checked by computing photometry
through a large aperture of $4.5$-pixel radius that encompassed the 
three stars in the 2011 and 2013 images after subtracting other nearby
sources---including the SN itself in the 2011 images. The resulting fluxes
were comparable (to within 20\%--30\%) with the summed SED in all
bands. However, due to the noise included in the large aperture, the
unknown aperture correction, and the arbitrary aperture centering,
this aperture photometry was too uncertain to be used in the 
following analysis. We instead computed
synthetic photometry using the A+B+C SED in the WFPC2 bands and
subtracted it from the 
observed SED of the pre-SN object. The resulting uncontaminated
photometry is shown in the right panel of Figure~\ref{fig:sed} and
listed in Table~\ref{tab:prph}. We note that the contamination accounted for 
about 34\%, 42\%, and 55\% of the pre-SN flux in the $F450W$, $F606W$,
and $F814W$ bands, respectively. The contribution in $F300W$ was
also calculated and it was subtracted from the upper flux limit of
the pre-SN object in that band. 

Figure~\ref{fig:sed} (right panel) also shows the SED of the remaining
object at the SN location in the 2011 images (at $t=1224$ days after
explosion), and the detection limits from the 2013 images (at $t=2063$
days after explosion). These calculations indicate that most of the
flux of the pre-existing object---even after removing stars A, B and C---has
disappeared. The SED at 1224 days may be mostly due to the fading SN,
as indicated by the strong decrease in flux at the $F438W$ and $F814W$
bands observed at 2063 days. The flux limits at 2063 days allow only for a small
contribution from any other object than the progenitor itself. Such
contributions are not greater than about 12\%, 8\% and 18\% of the
pre-SN flux in $F450W$, $F606W$, and $F814W$, respectively. The limits
in the bluest bands ($F275W$ and $F336W$) place strong constraints on
the near-UV emission of any putative left-over object (see
Section~\ref{sec:pre}). 

\begin{deluxetable}{lccc}  
\tabletypesize{\small} 
\tablecolumns{4} 
\tablewidth{0pt} 
\tablecaption{Fits to the SED of the putative progenitor and neighboring objects.\label{tab:sed}} 
\tablehead{ 
\colhead{Object} & \colhead{$T_{\mathrm{eff}}$} & \colhead{$L/L_\odot$}  & \colhead{Spectral} \\ 
\colhead{} & \colhead{(K)} & \colhead{} & \colhead{Type} 
}
\startdata 
Star A & $9180\pm700$  & $9400\pm1200$  & A1 Iab\\
Star B & $7180\pm190$  & $7640\pm210$   & F2 Iab\\
Star C & $4300\pm150$  & $8580\pm170$   & K3 Iab\\
 & & & \\
Progenitor\tablenotemark{a} & 7600--20000 & $2.6$--$20\times 10^4$ & B--A Ia
\enddata 
\tablenotetext{a}{One-sigma ranges are given because $T_{\mathrm{eff}}$ and
  $L/L_\odot$ are highly correlated as shown in Figure~\ref{fig:sedfit}.}
\end{deluxetable} 

\section{THE PROGENITOR OF SN~2008ax}
\label{sec:prog}

\subsection{Progenitor SED}
\label{sec:pre}

\begin{figure*}[htpb] 
\epsscale{1.0}
\plottwo{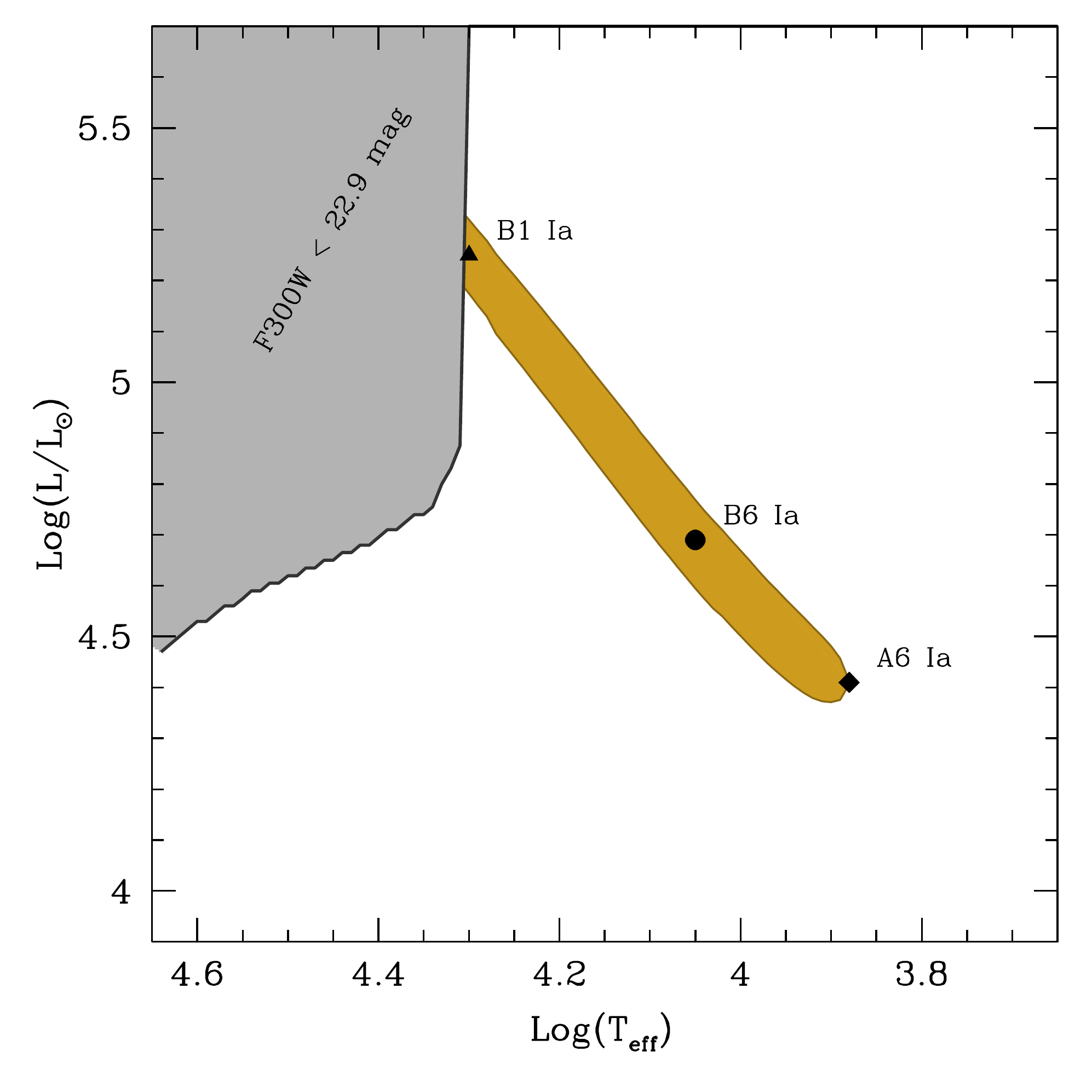}{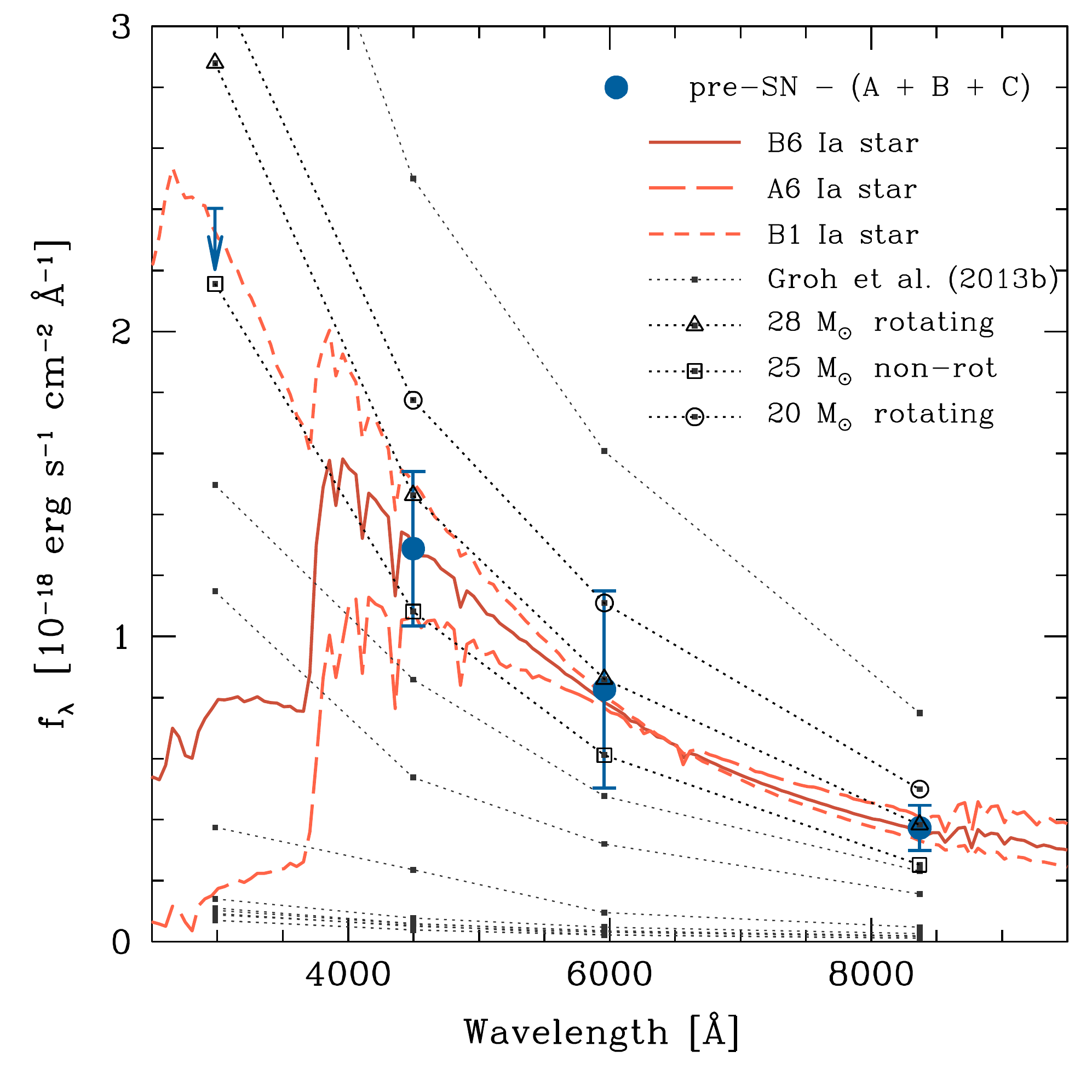}
\caption{Single-star progenitor fits. ({\em Left panel}) Confidence
  region with 68\% probability (yellow  
  shaded area) in the HRD from fits of Kurucz atmosphere models to the
  pre-SN object photometry. Red symbols show the parameters of model SED
  shown in the right panel, as labeled. The gray shaded region is
  forbidden by the detection limit in the $F300W$ band. 
  ({\em Right panel}) Uncontaminated photometry of the pre-SN
  object (blue circles and arrow) compared with model atmosphere SED
  (red lines). The best-fit model is shown with
  a solid line, rougly corresponding to a B6 Ia star. The long-dashed line
  shows the lowest-temperature model within the 68\% probability
  region, corresponding to an A6 Ia star. The highest-temperature
  model allowed by the detection limit in $F300W$ is shown with a
  short-dashed line and represents a B1 Ia star. Synthetic photometry
  from model atmospheres by \citet{Groh13b} are shown with gray
  squares connected by dotted lines. Three of these models that are
  described in Section~\ref{sec:pre} are highlighted: a 28-$M_\odot$
  rotating star (black triangles), a 25-$M_\odot$ non-rotating star
  (green squares), and a 20-$M_\odot$ rotating star (red circles).
  \label{fig:sedfit}}
\end{figure*}

\noindent We fit the revised photometry of the progenitor object after
removing the flux from the three nearby stars by assuming the SED of a
single star using the atmosphere models of \citet{Kurucz93}. The
results of such a fit are illustrated in Figure~\ref{fig:sedfit}. The
best-fit model is that with $T_{\mathrm{eff}}=11200$ K and
$\mathrm{Log}(L/L_\odot)=4.69$, roughly corresponding to a B6 Ia
star. The uncertainties in the photometry and lack of detection in the
$F300W$ band allow for a relatively wide range of solutions. As shown
by the confidence region in Figure~\ref{fig:sedfit}, temperatures can
be as low as $T_{\mathrm{eff}}=7600$ K. Solutions with
$T_{\mathrm{eff}} \geq 20000$ K fail to comply with the upper limit
in the $F300W$ band. Most importantly, the resulting temperatures and
luminosities are highly correlated. The solutions are within the realm
of bright supergiants, with spectral types between B1 and A6. The
stellar radii of the fitting models lie within a narrow range of
40--70 $R_\odot$. 

Alternatively, we tested the atmosphere models of SN progenitors
computed by \citet{Groh13b} for rotating and non-rotating massive stars
of solar metallicity. Their models predict that SN-IIb progenitors
should end their evolution as red supergiants, yellow hypergiants,
Wolf-Rayet or luminous blue variable (LBV) stars, depending on the initial
mass and rotation. In Figure~\ref{fig:sedfit}, right panel, we show
the SED of the models in \citet{Groh13b} computed from synthetic
photometry through the WFPC2 filters. Based on rotating models and the
uncorrected pre-SN photometry from \citet{Crockett08},
\citet{Georgy12} showed the progenitor of SN~2008ax was
well-reproduced by a 20-$M_\odot$ star initially rotating at 40\% of the
critical velocity. \citet{Groh13a} later reinterpreted the final stage
of that model as an LBV star. At the time such an LBV progenitor was
able to explain not only the available pre-SN photometry, but also it
had a relatively low mass to explain the SN light curve (see
Section~\ref{sec:hyd}), and a small amount of H to produce an
SN~IIb. With the revised distance and uncontaminated photometry
presented here, however, this model no longer fits the pre-SN
photometry, as shown in Figure~\ref{fig:sedfit}.  

A better fit to the revised pre-SN photometry in the $F450W$, $F606W$
and $F814W$ bands is achieved using the rotating model with 28
$M_\odot$ from \citet{Groh13b}. However, the UV emission is too bright
to comply with the upper limit  in the $F300W$ band (see 
Figure~\ref{fig:sedfit}). Another problem of the model is the low H
abundance at the final stage: the progenitor would be a WN star that
would produce a Type Ib SN. A further complication is the final
progenitor mass being $\approx 8$ $M_\odot$, which too large to
explain the light-curve timescale (Section~\ref{sec:hyd}). A possibly
better solution could be that of the non-rotating 25-$M_\odot$ model,
which according to \citet{Groh13b} leads to a final configuration that
is compatible with a SN-IIb event. As shown in the right panel of
Figure~\ref{fig:sedfit} (green squares), the model is slightly
underluminous in all bands. If, however, the distance is assumed to be
shorter within the uncertainty the fit to the observed pre-SN
photometry can be improved. As mentioned before, the final mass of the
model of $8.2$ $M_\odot$ poses a problem when one considers the shape
of the SN light curve. 

\begin{figure*}[htpb] 
\epsscale{1.0}
\plottwo{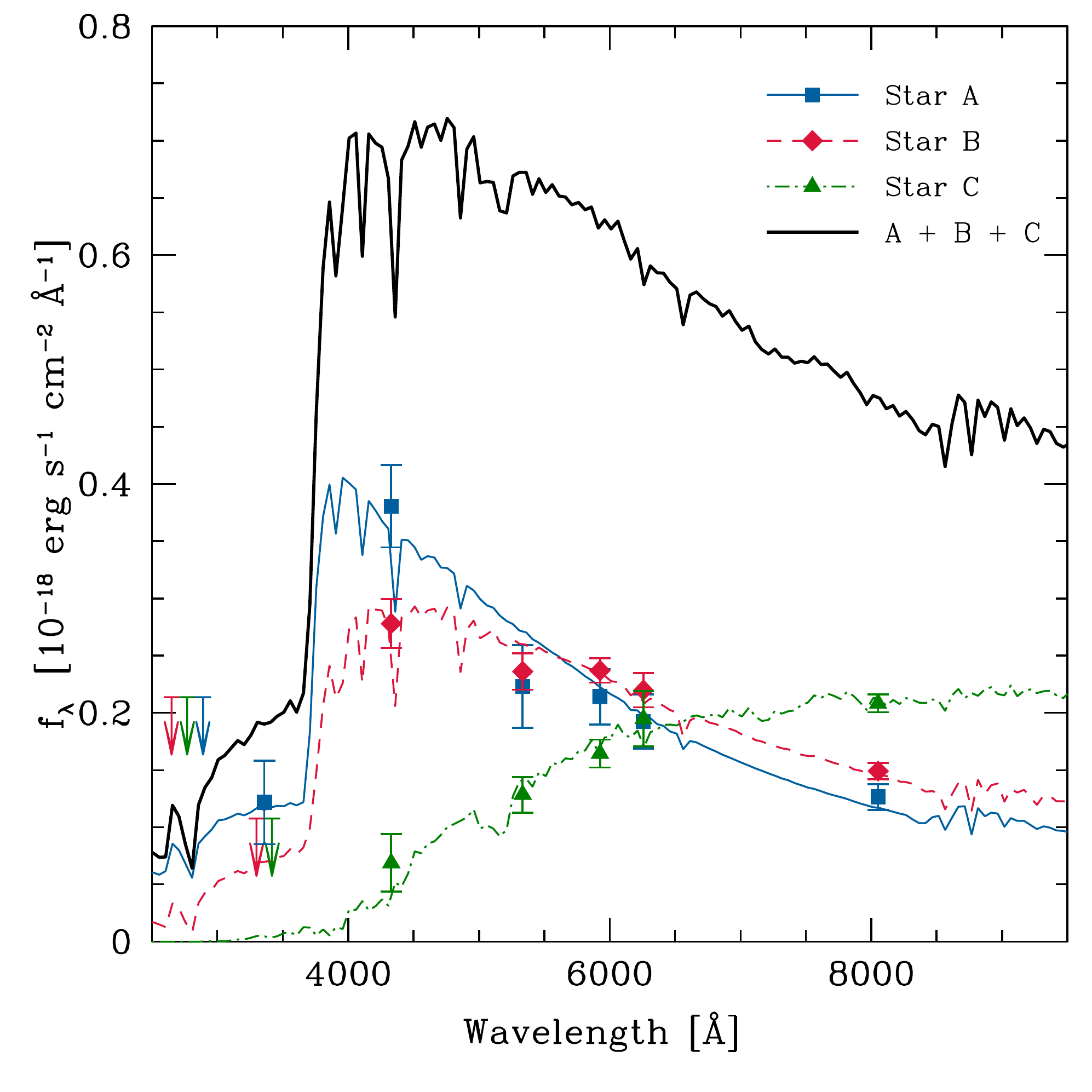}{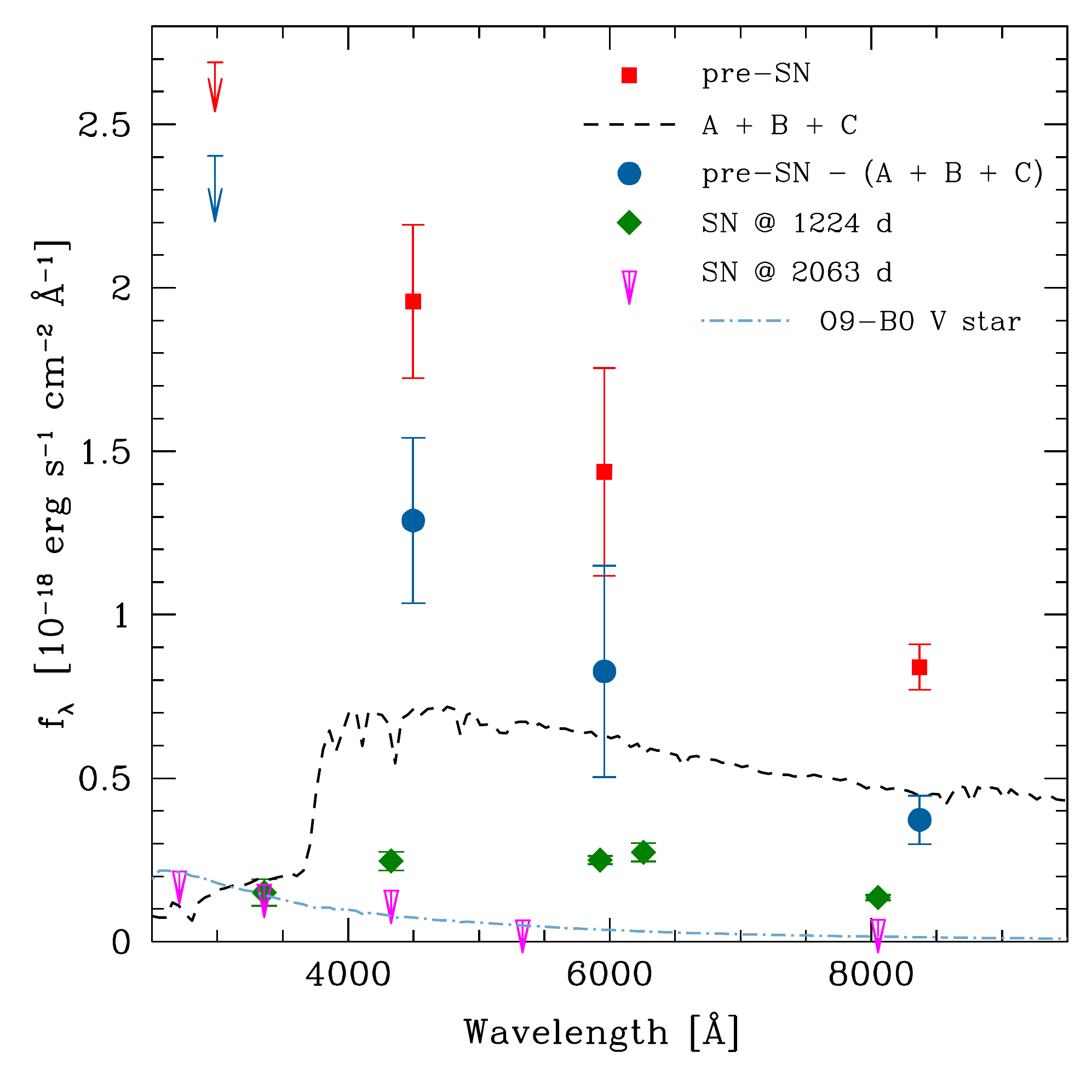}
\caption{Photometry and SED fits of objects in the SN site. ({\em Left
  panel}) Stars in the vicinity of SN~2008ax (A: blue squares; B: red
  diamonds; green triangles) with fitted stellar atmosphere SED from
  \citet{Kurucz93} (A: solid blue; B: dashed red; C: dot-dashed
  green). The arrows show upper limits in the $F275W$ and $F336W$
  bands. The black line is the sum of the fitted SED of the three
  stars. ({\em Right panel}) Pre-explosion photometry from WFPC2
  including stars A, B and C (red squares and red arrow for the upper
  limit in $F300W$), total SED of stars A, B and C (black dashed
  line), uncontaminated pre-SN photometry (blue circles and arrow),
  SN photometry from WFC3 obtained in 2011 at 1224 days
  after explosion (green diamonds), upper limits of SN flux from WFC3
  obtained in 2013 at 2063 days after explosion (magenta arrows). The
  blue dashed-dotted line shows the most-luminous allowed companion
  star in the case of a binary progenitor, corresponding to a O9--B0
  type main-sequence star.
  \label{fig:sed}}
\end{figure*}

A different scenario that appears to be recurrent among
stripped-envelope SNe and SNe~IIb in particular is an interacting
binary progenitor. In this case, the primary (donor) star explodes
after transferring most of its H-rich envelope to the secondary
(accreting) star. As a consequence, the secondary may remain after the
explosion. The detection limits from 2013 place constraints on the
brightness of any putative left-over companion star. The most luminous
companion star allowed is an O9-B0 type, main sequence star, as shown
on the right panel of Figure~\ref{fig:sed}. In Section~\ref{sec:hyd} we
provide support for a binary progenitor based on the mass of the
pre-SN object obtained from hydrodynamical modelling of the SN light
curve and expansion velocities. In Section~\ref{sec:bin} we
present a possible binary system computed with our stellar evolution
code that accounts for the pre-SN properties along with complying with
constraints from observations of the SN itself.

\subsection{Hydrodynamical Models}
\label{sec:hyd}

\noindent A set of explosion models were calculated in order to derive
the possible physical parameters of SN~2008ax. For this purpose we
used a one-dimensional Lagrangian LTE hydrodynamical code \citep{Bersten11}
applied to stellar evolution pre-SN structures \citep{Nomoto88} with
different He core masses. The methodology is the 
same that we applied in previous analyses of SNe~IIb \citep[e.g.,
  see][]{Bersten12,Bufano14}, i.e. we first derived 
global parameters focusing on the explosion of H-free structures to
fit the main peak of the bolometric light curve and the expansion
velocities, and then we used such parameters to analyze the effect on
the early light curve of a thin H-rich envelope with different
extensions. In order to compare models and observations on the same
time frame, we adopted an explosion date of $\mathrm{JD}=245 4528.80
\pm 0.15$ from \citet{Taubenberger11}. 

The bolometric light curve was derived using the color-based
bolometric corrections provided by \citet{Lyman14}. 
We adopted this approach over integrating the flux in the
observed bands to obtain a more homogeneous result at all epochs
independently of the wavelength coverage. The bolometric corrections
take into account contributions from the UV and IR ranges that can
be significant depending on the SN color. We employed 
the $BVRI$ and $g'r'i'$ light curves published by
\citet{Pastorello08} augmented by $BVRI$ observations of
\citet{Taubenberger11}. Bolometric corrections were derived using the
calibrations for $(B-V)$, $(B-R)$, $(B-I)$, $(V-R)$, $(V-I)$,
$(g'-r')$, and $(g'-i')$. The colors above were corrected by Milky-Way
and host-galaxy reddening as derived in
Section~\ref{sec:dist}. Observed magnitudes were corrected for
extinction and the resulting bolometric 
magnitudes were transformed to luminosities using the distance to
NGC~4490 also given in Section~\ref{sec:dist}. For each color a
different result was obtained, thus yielding seven bolometric light
curves. These were finally combined by interpolating
each light curve to the epochs of the others and averaging the results.

In order to approximate the photospheric velocities calculated by our
models, we adopted line expansion velocities derived from Doppler
shifts of the \ion{He}{1}~$\lambda$5876 and \ion{Fe}{2}~$\lambda$5169
absorptions. \ion{He}{1}~$\lambda$5876 velocities were taken from
\citet{Chornock11} and \citet{Taubenberger11}, while
\ion{Fe}{2}~$\lambda$5169 velocities were obtained from the latter
work. Roughly after day 10 since explosion 
\ion{He}{1}~$\lambda$5876 velocities began to flatten as the
photosphere receded beyond the bottom of the He-rich layer. At such
late times \ion{Fe}{2}~$\lambda$5169 presumably provides a closer
approximation of the photospheric velocity. At early times, the
\ion{Fe}{2}~$\lambda$5169 velocity has a flat behavior that may be due
to contamination from other lines as the \ion{Fe}{2} lines are rather
weak. We thus adopted the \ion{He}{1} velocities at $t\lesssim10$ days
and the \ion{Fe}{2} velocities afterward. 

\begin{figure*}[htpb] 
\epsscale{1.0}
\plottwo{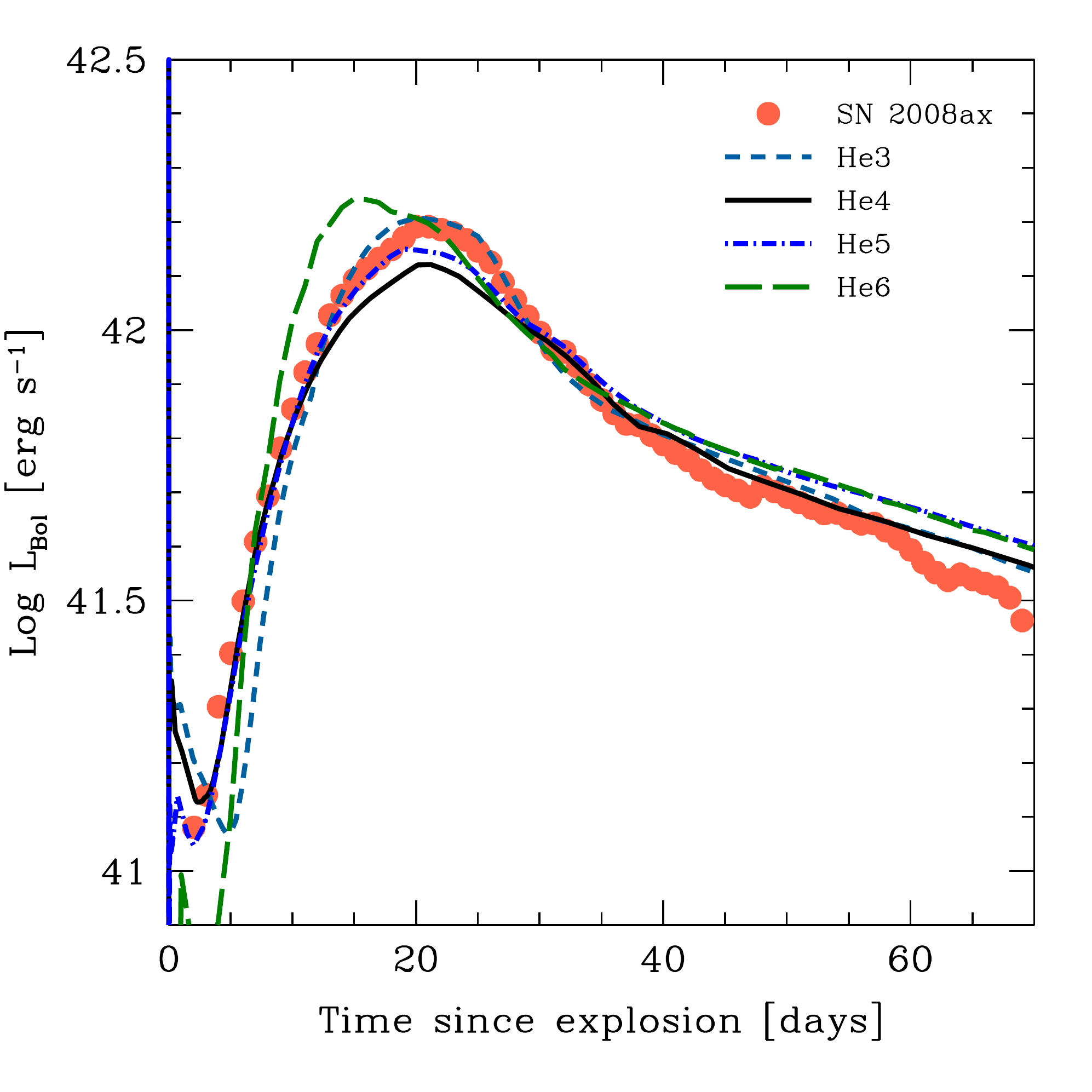}{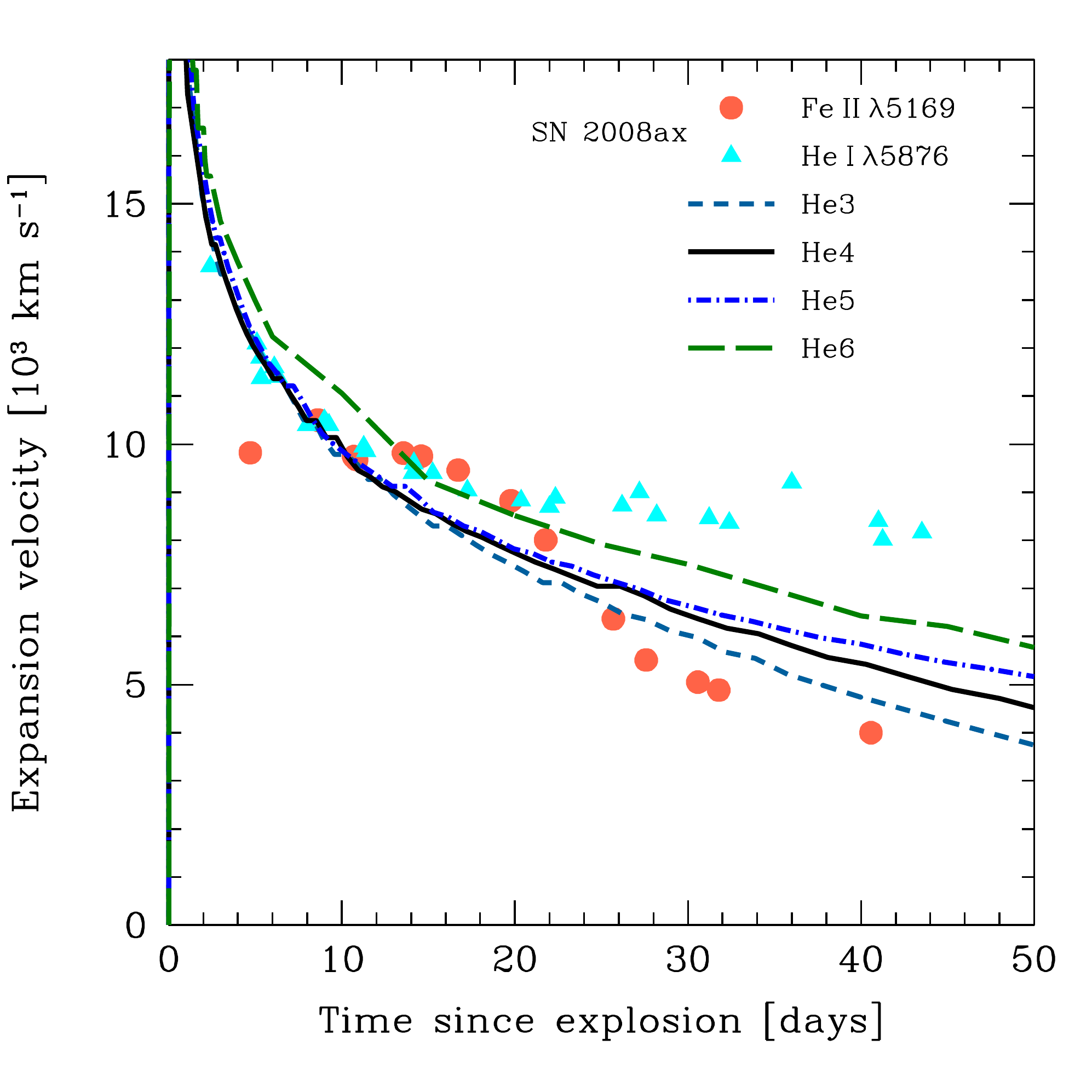}
\caption{\label{fig:hydmass} {\em (Left panel)} Observed bolometric light
  curve of SN~2008ax (circles) compared with the results of the light
  curve calculations for models He3 (dotted line), He4 
(solid line) and He5 (dashed line). {\em (Right panel)}
  Evolution of the photospheric velocity for models He3 (dashed
  line), He4 (solid line) and He5 (dash-dotted line) compared  
  with measured line velocities from the \ion{He}{1}~$\lambda$5876
  (triangles) and \ion{Fe}{2}~$\lambda$5169 (circles) lines.}        
\end{figure*}

In the first stage of the modeling we derive the ejected mass,
explosion energy and radioactive nickel production by fitting the
overall shape of the bolometric light curve and expansion velocity
evolution. Figure~\ref{fig:hydmass} shows the observations compared
with four models of different progenitor masses: (1) model He3 with He
core mass of $3.3$ $M_\odot$, explosion energy of $8 \times 10^{50}$
erg, and $^{56}$Ni mass of $0.07$ $M_\odot$, (2) model He4 with He
core mass of 4 $M_\odot$, explosion energy of $1 \times 10^{51}$ erg,
and $^{56}$Ni mass of $0.05$ $M_\odot$, (3) model He5 with He core mass
of 5 $M_\odot$, explosion energy of $1.2 \times 10^{51}$ erg, and
$^{56}$Ni mass of $0.06$ $M_\odot$, and (4) model He6 with He core
mass of 6 $M_\odot$, explosion energy of $2 \times 10^{51}$ erg, and
$^{56}$Ni mass of $0.07$ $M_\odot$.

We consider that models He3, He4, and He5 give a fair representation
of the observations judging all the uncertainties involved, whereas
model He6 provides an overall worse fit to the light curve shape and
velocity evolution. The range of validity of the model parameters  
is roughly given by the three former solutions. Note that 
models He4 and He5 provide a better fit on the rising part of the
bolometric light curve than model He3, but the latter gives a better
match to the main peak and subsequent decline. While He5 produces a slightly
overestimation of the tail luminosity, He4 has a lower peak and
slightly wider light curve than what was observed. On the other hand,
He3 gives the best approximation to the late-time photospheric velocity
evolution as traced by the \ion{Fe}{2}~$\lambda$5169 line. Even if 
models He4 and He5 provide higher photospheric velocities than
the observed ones we consider that they are acceptable given the
uncertainties involved. Contrarily, model He6 departs from the
observations in an unsolvable manner. Its light curve is too 
and wide, the initial rise is not well-reproduced, and the late-time
velocities are too large. If the energy were lowered to improve the
match to the velocities, the resulting light curve would be even wider
and the initial luminosity evolution would be even lower.

This analysis suggests that the
progenitor of SN~2008ax had a He core mass of $\lesssim 5$ $M_\odot$. In
order to have this low He core mass, the progenitor necessarily had a
zero-age main sequence (ZAMS) mass below 25 $M_\odot$, thus favoring
a binary origin. Our conclusions are consistent with previous results
based on analytical modeling \citep{Pastorello08,Roming09,Taubenberger11}.
  
Early photometry obtained within a few days from explosion can serve
to reveal the progenitor structure and in particular, the extent of its
H-rich envelope \citep[e.g., see][]{Bersten12,Bersten14}. During this
phase, before radioactivity begins to dominate, the light curve is
regulated by the physical conditions of the shock-heated ejecta. This in
turn leads to an initial luminosity decline usually referred to as the
``cooling phase''. The more extended the progenitor structure, the
higher the luminosity and temperature of the emitting region right
after shock breakout, and the slower the subsequent decline. The effect on
temperature makes the cooling phase more noticeable in the blue bands
than in the red ones.

Interestingly, SN~2008ax was observed in the UV range by the {\em
  Swift} satellite soon after the explosion \citep{Roming09}. The {\em
  Swift}/UVOT data in the UV bands showed an initial decline lasting for at
least four days after explosion. This decline was not observed in the
optical range, although a slower initial rise may be seen in the {\em
  Swift} $b$ band \citep{Roming09} and in the $g'$ band
\citep{Pastorello08}. These observations are reminiscent of a post
shock cooling phase. We thus investigated whether the early light
curves provided information about the progenitor extent based on our
hydrodynamical modelling.

\begin{figure*}[htpb] 
\epsscale{1.0}
\plottwo{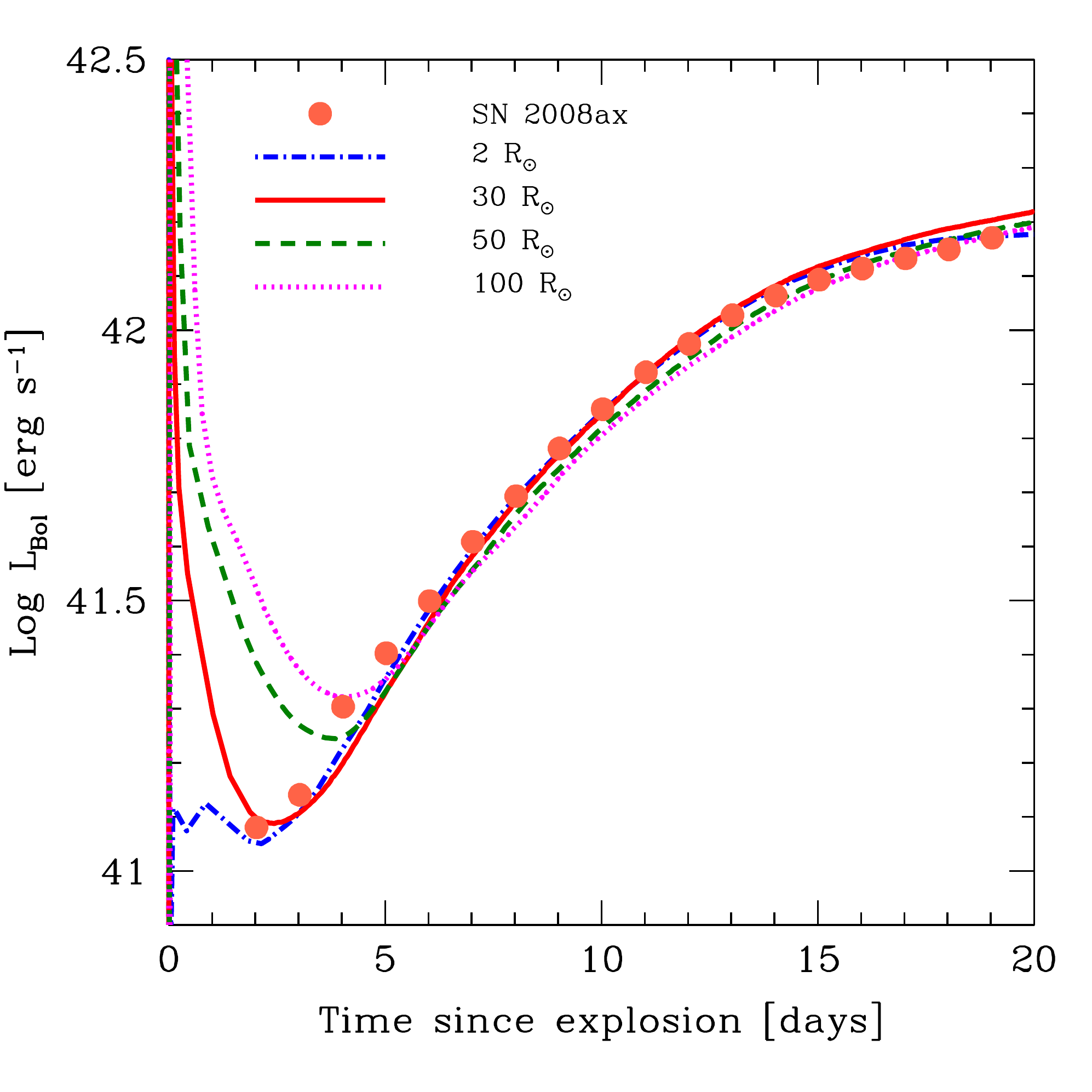}{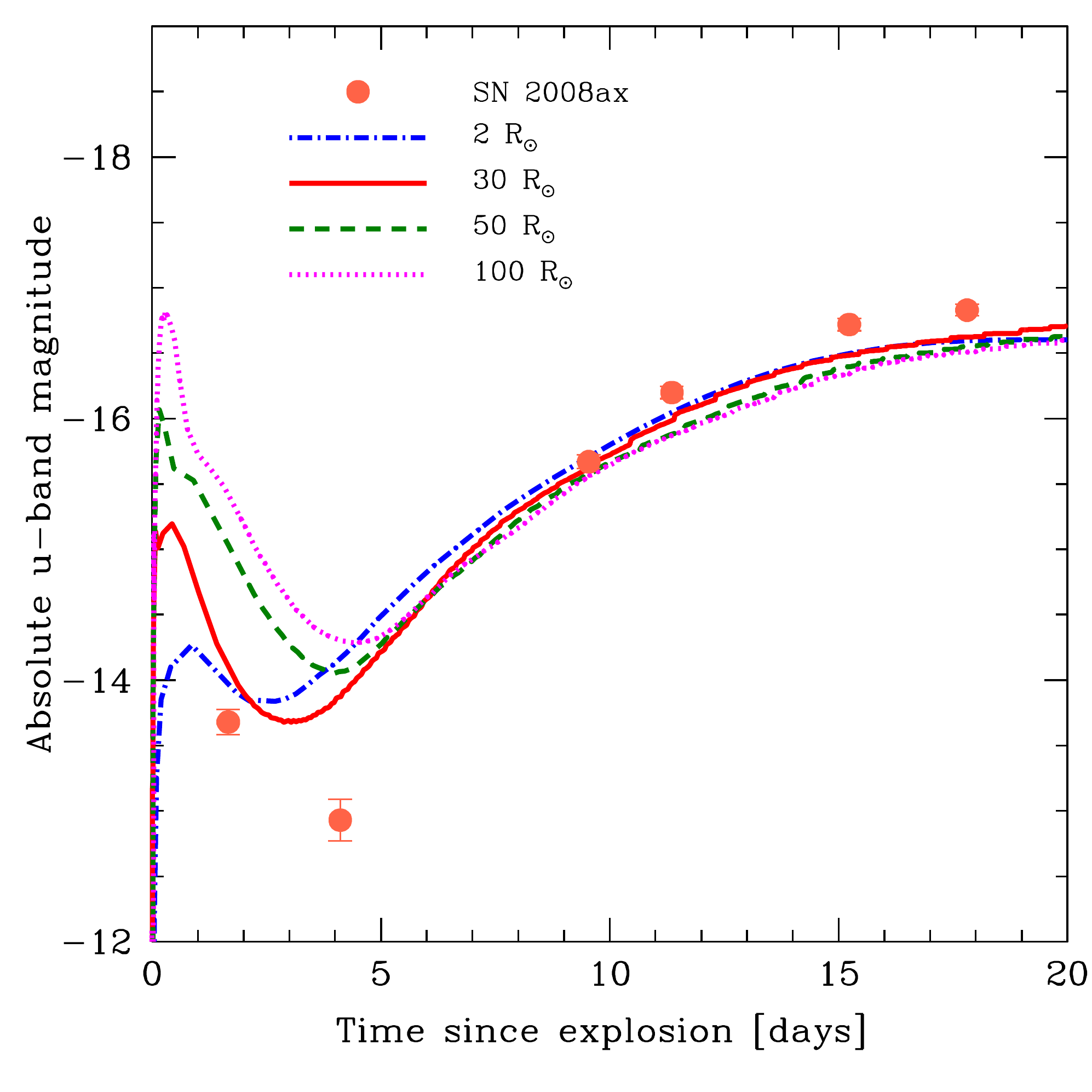}
\caption{\label{fig:hydrad} {\em (Left panel)} Observed bolometric light
  curve of SN~2008ax (circles) at early times compared with
  hydrodynamical models for progenitors with 5 $M_\odot$ helium cores
  and different extensions of the H-rich envelopes (lines). Model He5
  from Figure~\ref{fig:hydmass} is that with 2 $R_\odot$.
  {\em (Right panel)} Observed {\em Swift}/UVOT
  $u$-band light curve \citep[circles;][]{Roming09} compared with the
  same models as in the left panel. A black body was assumed to
  calculate the $u$-band luminosity from the models. 
  }        
\end{figure*}

We considered our H-free He5 initial stellar structure, which has a
radius of $R=2$ $R_\odot$, and modified it by adding thin H-rich
envelopes of different extent to $R=30$, 50, and 100 $R_\odot$ and
masses of $0.02$--$0.04$ $M_\odot$. With
these initial structures we computed the hydrodynamics using the same
explosion parameters as in the model shown in
Figure~\ref{fig:hydmass}. The left panel of Figure~\ref{fig:hydrad}
shows the comparison of the resulting bolometric light curves with the
observations at early times. Clearly, as the radius increases the
initial decline becomes more evident. After about five days since
explosion, all the models converge as the light curve becomes
regulated by $^{56}$Ni radioactivity. The comparison suggests that
progenitor radii up to about 30 $R_\odot$ are allowed. 

A caveat should be made about the accuracy of the bolometric
luminosity at the earliest times when a significant fraction of the flux
is expected to be emitted in the UV range. According to
\citet{Pritchard14}, the observed colors of SN~2008ax indicate that
the fraction of the total flux included in the {\em Swift} UV range
is well below 10\% at all epochs. And even if the bolometric
corrections from \citet{Lyman14} can account for this contribution 
(see their Section 5.1), any extra flux at shorter wavelengths may have
been missed. This is why in the following we decided to directly
compare our models with light curves in individual {\em Swift}/UVOT
bands. 

For that purpose we assumed a black body emission to 
convert the modeled bolometric light curves into broad-band light
curves. This is a simple assumption, although at early times it may
provide a valid approximation as the ejecta are sufficiently dense. We
constrained our analysis to the $u$-band data since it showed the most
clear initial decline. Bluer UV bands showed only upper limits at this
early stage. The right panel of Figure~\ref{fig:hydrad} shows the
comparison of our extended He5 models with the $u$-band light
curve. Absolute magnitudes were obtained from the observations by
assuming a distance modulus and extinction as derived in
Section~\ref{sec:dist} (i.e., $u$-band extinction of $A_u=1.47$
mag). Although the models fail to exactly reproduce the low luminosity
at $t < 5$ days, presumably due 
to the black-body approximation, we can qualitatively see that the
more extended the progenitor the larger the initial decline and the
later the light-curve minimum occurs. Also the extended models produce a
smaller contrast between the minimum and the subsequent maximum of the
light curve. The apparent duration and depth of the observed minimum
would indicate that the most favored models are those between 30 and
50 $R_\odot$. 

We performed a similar analysis based on the He3 and He4 models and
found a worse comparison with the observations. In particular, as
the core mass decreases, the $u$-band light curve becomes more
luminous. This is a consequence of the increased temperatures reached
by the ejecta due to the shock wave propagation. Although the 
approximations involved may prevent a definitive conclusion to be
driven, the overall trends of the $u$-band analysis seem to indicate
that the mass of the progenitor was between 4 and 5 $M_\odot$ and its
radius was between 30 and 50 $R_\odot$. Indeed, the location of the
pre-SN object in the Hertzsprung-Russell diagram (HRD; see
Section~\ref{sec:bin}) indicates an extended structure for the
progenitor of $R \approx 40$ $R_\odot$.  

\subsection{A Binary Progenitor}
\label{sec:bin}

\noindent We now consider whether the results from the
previous sections can be interpreted in the frame of the theory of binary
stellar evolution. For this purpose we computed  
several models with the aim of accounting for the pre-supernova
photometry and, simultaneously, the observed light curve of
SN~2008ax. A successful model should produce a progenitor with a final
mass of $\approx 4 M_{\odot}$ to reproduce the SN light curve, and a
small amount of hydrogen in the outermost layers to satisfy the SN
classification as Type~IIb.

We employed the code described in \citet{Benvenuto03}, updated to
consider the case of massive binaries as presented in
\citet{Benvenuto13}. Our stellar code handles the mass
transfer process in a fully implicit way. We considered models with
solar chemical composition and ignored the effects of
rotation. Furthermore, we assumed that the companion star is able to
retain a fraction $\beta$ of the material transferred by the primary
(the one that explodes). $\beta$ is kept fixed throughout the entire
evolution of the pair. We ended the calculations at oxygen
exhaustion. Since the remainder of the primary evolution occurs during a
very short time scale and only affecting its core, it is safe to assume
that the stars do not significantly move any further in the HRD.

We explored the parameters of the problem---the masses of the
components and the initial orbital period, $P_{\mathrm{orb}}^{i}$---based on
our study of SN~2011dh \citep{Benvenuto13}. In comparison, for
SN~2008ax we found slightly larger initial masses and a much shorter
$P_{\mathrm{orb}}^{i}$. The observations of SN~2008ax are 
fulfilled with a pair of 18$M_{\odot}$+12$M_{\odot}$ on an orbit with
$P_{\mathrm{orb}}^{i}=5$ days, while for SN~2011dh we proposed a pair with
16$M_{\odot}$+10$M_{\odot}$ and $P_{\mathrm{orb}}^{i}= 125$ days.  

\begin{figure*}[htpb] 
\begin{center}
\includegraphics[scale=0.50,angle=270]{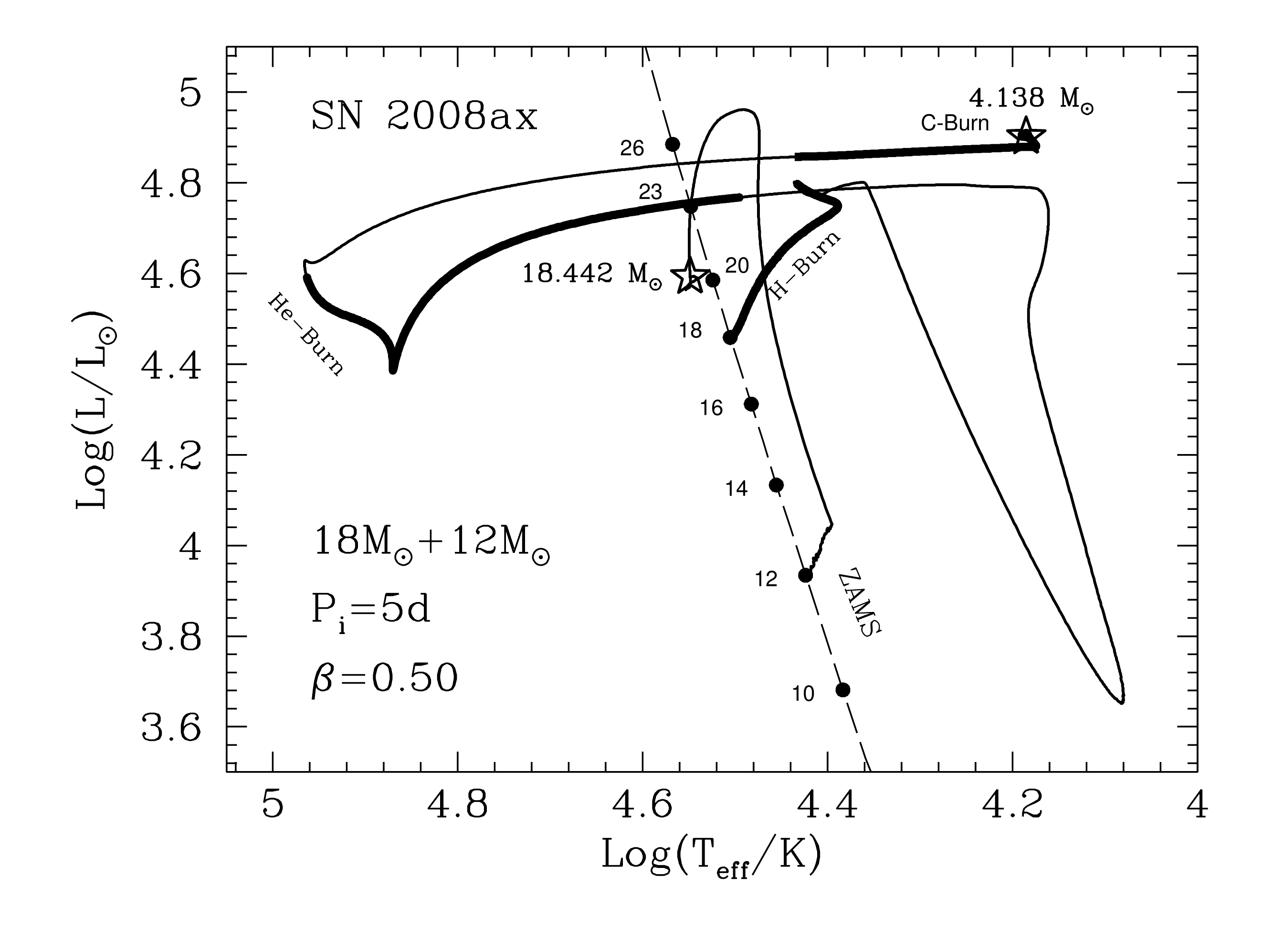}
\caption{\label{fig:HRD} Evolutionary tracks in the
Hertzsprung-Russell diagram for a binary system of
18$M_{\odot}$+12$M_{\odot}$ on an orbit with $P_{\mathrm{orb}}^{i}=5$ 
days. The ZAMS locus is shown for reference by a dashed line
with some mass values (in solar units) denoted with filled dots. Each
track starts at the corresponding location on the ZAMS. Thick portions
of the primary track indicate stages of active core burning. The
conditions at the moment of explosion are shown with stars. The
primary star undergoes RLOF after hydrogen core 
exhaustion. When it detaches it has a mass of $4.376$ $M_{\odot}$ and
starts a blueward loop on which it burns out its helium core. Since
then it swells and undergoes a final RLOF on which it completes
carbon and subsequent nuclear cycles (neon, oxygen, and silicon
burning), and it finally explodes.  
Simultaneously, the companion star accretes half (because we set
$\beta= 0.50$) of the transferred material, reaching a final mass of
$18.442$ $M_{\odot}$.} 
\end{center}
\end{figure*}

In Figure~\ref{fig:HRD} we show the resulting evolutionary track for
the case of $\beta= 0.50$. Shortly after the end of the core hydrogen
burning stage, the primary star undergoes Roche lobe overflow (RLOF), i.e., 
a Class B mass transfer episode. The main mass transfer episode starts
at an age of $9.785$ Myr and lasts for only 26 Kyr. The star detaches from
its Roche lobe with a mass of $4.376$ $M_{\odot}$; thus, the mean mass
transfer rate is of $5.08 \times 10^{-4}$ $M_{\odot}$ yr$^{-1}$ while the
maximum value is of $2.30 \times 10^{-3}$ $M_{\odot}$ yr$^{-1}$. Core helium
burning starts soon after the end of the RLOF, and most of it is
burned during the large blue loop in the HRD. When central
helium is exhausted the star begins to swell again, evolving to lower
effective temperature conditions. Then, the primary star begins to
burn carbon and quickly undergoes the second RLOF that lasts
up to explosion. The companion star reacts to accretion by swelling
appreciably, but it does not fill its own Roche lobe at any moment;
therefore, the binary never reaches a contact configuration and thus
avoids suffering a common-envelope episode. 

The primary explodes at an age of $11.02$ Myr. At these conditions, the
pre-SN has a mass of $4.14$ $M_{\odot}$ and a radius of
$40.7$ $R_{\odot}$. Most of its radius is occupied by the outer hydrogen
rich layer (outward of $1.4$ $R_{\odot}$) containing only
$0.06$ $M_{\odot}$. The total amount hydrogen in the outer layer is of
$\approx 3.6\times10^{-3}$ $M_{\odot}$. At that moment, the companion
has a mass of $18.44$ $M_{\odot}$ with a luminosity and effective
temperature very similar to those corresponding to a 20 $M_{\odot}$
star on the Zero Age Main Sequence (ZAMS, see Figure~\ref{fig:HRD});
so, it is somewhat overluminous.  

The parameters chosen for the binary system satisfy the
effective temperatures and luminosities for both stars just before the
explosion, and also a primary structure capable of accounting the
light curve and classification of SN~2008ax. However, we should 
remark that this is not a unique solution because the parameters
  are degenerate. For example, we may have chosen a higher
initial mass for the companion and a lower value of $\beta$ to reach a
similar final mass. In any case, it seems virtually impossible to find
a final mass of the primary of 4--5 $M_\odot$---suitable to account for the
observed light curve---if the initial primary mass is modified by more
than about $\pm 2$ $M_{\odot}$. 

The results presented in this section allow us to state that
binary stellar evolution is able to provide a successful progenitor
scenario to explain the main characteristics of SN~2008ax.

\newpage
\section{CONCLUSIONS}
\label{sec:concl}

\noindent SN~2008ax has been one of the best-studied SNe~IIb to
date. Along with extensive, multi-wavelength follow-up observations,
the existence of 
deep pre-explosion {\em HST} images placed this object among a
selected group of SNe~IIb for which a progenitor could be
characterized, as first done by \citet{Crockett08}. Since then,
additional observations of the SN site were obtained with even higher
resolution using the WFC3 camera on board {\em HST}. The new images
allowed us to revisit the progenitor candidate and to discover that
it was in fact a multiple source. After subtracting the light from
three nearby stars, the revised progenitor photometry resulted
significantly fainter and bluer than what was measured on the
pre-explosion images. Moreover, the new images showed a fading source
coincident with the SN location and demonstrated that most of the
light from the pre-SN source had disappeared. This observation
unambiguously showed the correct identification of the exploding
star. SN~2008ax is thus the third SN~IIb, after SNe~1993J and 2011dh,
with a firm progenitor detection.

Comparing the revised photometry with stellar atmosphere models, we
found that the progenitor, if single, was compatible with a B to mid-A
type supergiant star. Contrarily, pre-SN models of WR and LBV stars
calculated by \citet{Groh13b}, some of which had been proposed for
SN~2008ax based on the contaminated photometry, are not compatible
with the new photometry. Detection limits from the 2013 images
allow for a relatively faint source to be left at the SN site. We
estimated that any such remaining object could be as luminous as an
O9--B0 main-sequence star.

Using hydrodynamical models to reproduce the bolometric light curve
and photospheric velocity evolution we arrived at the conclusion that
the exploding object had a relatively low mass of 4--5 $M_\odot$,
similarly to what had been concluded based on other techniques. From a
comparison with the early light curves, especially those in the UV
range, we additionally concluded that the progenitor must have had a
relatively extended structure, with a low-density H-rich envelope
reaching about 30--40 $R_\odot$. Such a low mass and extended
radius disfavor a massive progenitor that loses the outer envelope
via stellar winds and ends its life as a WR or LBV star. We note that
even if the favored progenitor appears as a blue supergiant star,
resembling those of SN~1987A-like events, the stellar
structure is substantially different in the case of SN~2008ax, with a
much lower-mass, extended H-rich envelope. The difference in H mass is
what explains the dissimilar light curve shape and spectroscopic
properties between the SN~IIb 2008ax and the SN~1987A-like objects. 

A viable alternative is that of an interacting binary
system. In this case, the progenitor is allowed to be less massive and
the envelope is lost through mass transfer to the companion star. We
were able to provide such a binary model using our stellar evolution
code. The model assumed a pair of 18$M_\odot$+12$M_\odot$ in a close
orbit (with initial period of 5 days) that naturally explained the
explosion of a nearly H-free, $\approx$4 $M_\odot$ star of $R \approx
40$ $R_\odot$ with a remaining main-sequence companion that complied
with the latest detection limits. The solution is not unique but it
demonstrates that the scenario is plausible. A definitive confirmation
and characterization of the proposed binary progenitor would require a
deep search for the companion star in future observations from space.

The case of SN~2008ax sheds light on the origin of SNe~IIb, which has
implications for the progenitors of core-collapse SNe in general. To
date there are three SNe~IIb with identified progenitors, namely
SNe~1993J, 2008ax, 2011dh, and one candidate for
SN~2013df. Remarkably, in all of these cases the progenitors have been
suggested to belong to binary systems. For 
SNe~1993J and 2011dh there is further evidence of the presence of a
hot companion star \citep{Maund04,Fox14,Folatelli14}. This fact points
to a possible common origin for SNe~IIb. 

The history of mass loss for SN-IIb progenitors is a crucial aspect to
understand their nature. From radio and X-ray observations and by
studying the late-time optical spectra and light curves, evidence of
substantial amounts of CSM has been found for SNe~1993J and
2013df. Contrarily, SNe~2008ax and 2011dh appear to have occurred in much
``cleaner'' environments. Such differences may indicate a different 
mass-loss history. In the interacting binary scenario, this may in
turn be linked with the strength and timing of the mass-transfer
episodes and with the ability of the accreting star to retain the
material. As pointed out by \citet{Maeda15}, the cases with denser CSM
correspond to more extended progenitors, of over 500 $R_\odot$. With a
progenitor radius of only $\approx$40 $R_\odot$ as derived from our
models, SN~2008ax would agree with the same picture. More objects are
necessary to confirm if this is a general rule for SNe~IIb and whether
the progenitor radius and CSM density follow a continuum of properties
or if they group into two distinct classes. Finally, the physical
reason for such a possible distinction is yet to be known and would
require detailed modeling of the progenitor evolution.

\acknowledgments 
We would like to thank the Hubble Legacy Archive Hotseat Office, in
particular Rick White, for valuable help with handling the
WFPC2/$F606W$ images.
This research has been supported by the World Premier International Research
Center Initiative (WPI Initiative), MEXT, Japan, and
by Grants-in-Aid for Scientific Research (3224004, 26400222, and
26800100). Support for HK is provided 
by the Ministry of Economy, Development, and Tourism's Millennium
Science Initiative through grant IC120009, awarded to The Millennium
Institute of Astrophysics, MAS. 
HK acknowledges support by CONICYT through FONDECYT grant 3140563.
The present work is based on
observations made with the NASA/ESA {\em Hubble Space Telescope}, 
and obtained from the Hubble Legacy Archive, which is a
collaboration between the Space Telescope Science Institute
(STScI/NASA), the Space Telescope European Coordinating Facility
(ST-ECF/ESA) and the Canadian Astronomy Data Centre
(CADC/NRC/CSA). Gemini data were provided by the Canadian Astronomy Data
Centre operated by the National Research Council of Canada with the
support of the Canadian Space Agency. We have made use 
of the NASA/IPAC Extragalactic Database (NED) which is operated by the
Jet Propulsion Laboratory, California Institute of Technology, under
contract with the National Aeronautics and Space
Administration. 


\end{document}